\documentclass[conference]{IEEEtran}
\IEEEoverridecommandlockouts

\usepackage{cite}
\usepackage{amsmath,amssymb,amsfonts}
\usepackage{algorithmic}
\usepackage{graphicx}
\usepackage{textcomp}
\usepackage{xcolor}
\usepackage{booktabs}

\usepackage{ifthen}

\newcommand{\bX}{\boldsymbol{X}}
\newcommand{\bx}{\boldsymbol{x}}
\newcommand{\bF}{\boldsymbol{F}}
\newcommand{\bb}{\boldsymbol{b}}
\newcommand{\bC}{\boldsymbol{C}}
\newcommand{\bc}{\boldsymbol{c}}

\newcommand{\bff}{\boldsymbol{f}}
\newcommand{\bS}{S}

\newcommand{\myparagraph}[1]{\vspace{0.3\baselineskip}\noindent{\textbf{#1.}}~}

\newcommand{\CDKM}{\texttt{CDKM}\xspace}

\newcommand{\Lloyd}{\texttt{Lloyd}\xspace}
\newcommand{\DPFMPS}{\texttt{DPFMPS}\xspace}
\newcommand{\VFLC}{\texttt{VFLC}\xspace}

\newcommand{\TriCD}{\texttt{TriCD}\xspace}

\newcommand{\FedTriCD}{\texttt{FedTriCD}\xspace}
\newcommand{\FFFKM}{\texttt{F$^3$KM}\xspace}
\newcommand{\Hamerly}{\texttt{Hamerly}\xspace}
\newcommand{\Yinyang}{\texttt{Yinyang}\xspace}
\newcommand{\ballkmeans}{\texttt{Ball $k$-means}\xspace}
\newcommand{\LBKC}{\texttt{LBKC}\xspace}
\newcommand{\Elkan}{\texttt{Elkan}\xspace}

\newcommand{\YearPredictionMSD}{\textsf{YearPrediction}\xspace}
\newcommand{\Epileptic}{\textsf{Epileptic}\xspace}
\newcommand{\NYC}{\textsf{NYC}\xspace}
\newcommand{\Kegg}{\textsf{Kegg}\xspace}
\newcommand{\Crime}{\textsf{Crime}\xspace}
\newcommand{\Shuttle}{\textsf{Shuttle}\xspace}
\newcommand{\Letter}{\textsf{Letter}\xspace}
\newcommand{\Mnist}{\textsf{Mnist}\xspace}
\newcommand{\Cifarten}{\textsf{Cifar10}\xspace}
\newcommand{\Cifarhundred}{\textsf{Cifar100}\xspace}

\usepackage{amsthm}
\usepackage{tikz}

\newtheoremstyle{mydefinition} 
  {1pt} 
  {1pt} 
  {\itshape} 
  {} 
  {\bfseries} 
  {.} 
  { } 
  {} 

\theoremstyle{mydefinition}

\newtheorem{theorem}{Theorem}

\newtheorem{lemma}{Lemma}

\usepackage{mathtools}

\usepackage{amsmath}
\usepackage{amsthm}
\usepackage{enumitem}
\usepackage{multirow}

\usepackage{xcolor}

\usepackage{ragged2e}
\usepackage{etoolbox}
\usepackage{xcolor}
\usepackage{algorithmic}
\usepackage[linesnumbered,ruled,vlined,boxed,noend]{algorithm2e}
\SetAlgoSkip{}

\usepackage{graphicx}
\usepackage[caption=false,font=normalsize,labelfont=sf,textfont=sf]{subfig}

\usepackage[
  colorlinks=true,
  linkcolor=black,
  citecolor=blue,
  urlcolor=blue
]{hyperref}

\begin{document}

\title{Highly Efficient Vertical Federated Clustering with Triangle-Inequality-based Pruning}

\author{
\IEEEauthorblockN{
Feiteng Nie\textsuperscript{1},
Shengkun Zhu\textsuperscript{1},
Zepeng Liu\textsuperscript{1},
Sheng Wang\textsuperscript{1*},
Yuan Sun\textsuperscript{2},
Zhiyong Peng\textsuperscript{1}
}

\IEEEauthorblockA{
\textsuperscript{1}School of Computer Science, Wuhan University,
\textsuperscript{2}La Trobe Business School, La Trobe University
}

\IEEEauthorblockA{
\{niefeiteng, whuzsk66, liuzp\_063, swangcs, peng\}@whu.edu.cn,
yuan.sun@latrobe.edu.au
}

\thanks{\textsuperscript{*} Sheng Wang is the corresponding author.}
}





\maketitle

\begin{abstract}
In many real-world scenarios, multiple parties hold complementary feature subsets for the same samples but cannot share raw data. 
This makes efficient clustering over distributed feature subsets an important and challenging problem.
While vertical federated learning (VFL) enables such collaboration, existing VFL clustering methods often require many communication rounds or substantial local computation.
To address these limitations, we first propose \TriCD, an exact pruning method for \CDKM that preserves the same cluster assignments under the same initialization.
\TriCD constructs triangle-inequality bounds on demand to skip redundant objective evaluations with negligible extra space.
Building upon this, we propose \FedTriCD, an efficient federated clustering algorithm.
By decomposing the global objective into client-local subproblems, \FedTriCD enables pruning-based local optimization using only local data.
\FedTriCD further reduces communication by allowing multiple local updates between server synchronizations.
A proximal regularization term is introduced to control the drift caused by infrequent communication.
Our theoretical analysis establishes the exactness and finite convergence of \TriCD and provides a bounded-error server-side update guarantee for \FedTriCD.
Experiments on multiple real-world datasets demonstrate the efficiency and clustering quality of the proposed methods.
\TriCD achieves up to a 50.07$\times$ speedup over \Lloyd's algorithm. 
Moreover, \FedTriCD achieves competitive clustering quality while requiring fewer local evaluations and fewer communication rounds than state-of-the-art federated clustering methods.
\end{abstract}

\section{Introduction}
Clustering is a fundamental task in unsupervised learning that automatically reveals latent structure by grouping similar samples in large-scale data.
Among clustering methods, $k$-means remains one of the most widely used due to its simple objective, scalability, and broad applicability in healthcare~\cite{gonen2014localized, Huang2022OVQA, John2015k-Shape, Tan2022MetaCare}, finance~\cite{chandramouli2006database, Arash2020Vertica, Zhou2009Graph}, and education~\cite{He2022Unsupervised, ZhangW0P23}.
As datasets continue to grow rapidly in both scale and dimensionality, $k$-means can become costly, making efficiency a central concern in practical deployment.
Fast $k$-means has therefore become an important research direction for improving the practicality of clustering~\cite{Elkan03Using, Hamerly10Making, ding15Yinyang,wang20Efficiency, zhang2024speeding,ji2025simplifying}.

\begin{figure}[t]
  \centering
  \includegraphics[width=\linewidth]{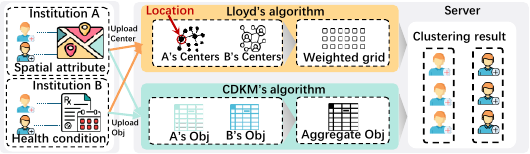}
  \vspace{-1.4em}
  \caption{Lloyd-style methods exchange local centers or center-derived summaries, whereas our method aggregates local objective values instead.}
  \label{fig:vfl_motivation}
  \vspace{-0.8em}
\end{figure}

In many real-world applications, features of the same samples are vertically partitioned across institutions, while legal and regulatory constraints prevent the sharing of raw data~\cite {GDPR2018, HIPAA1996}.
Consider a cross-institution chronic disease management example in Fig.~\ref{fig:vfl_motivation}. 
Institution A holds patients' diagnosis and treatment records, whereas Institution B maintains their post-discharge follow-up information. 
Joint clustering over these complementary features can help identify patient groups with similar care needs and support group-specific treatment strategies. 
However, existing methods do not simultaneously satisfy the efficiency and information-sharing constraints of this setting.
Most fast $k$-means techniques assume centralized access to complete features and cluster statistics, making their pruning rules difficult to apply directly under vertical feature partitioning.
Moreover, their bound-based pruning mechanisms often require retaining and updating point-to-centroid bounds, introducing substantial additional storage and maintenance overhead.
Meanwhile, existing Lloyd-based VFL~\cite{fu2023feast, zhang2026rap} clustering methods~\cite{Ding16KMeans, Yang19Federated, liu2024vertical, Fu2022BlindFL, kou2026fedharmony} often exchange center-derived summaries, which can reveal institution-specific prototypes and sensitive feature patterns, as illustrated in Fig.~\ref{fig:vfl_motivation}.

\begin{figure}[t]
  \centering
  \includegraphics[width=\linewidth]{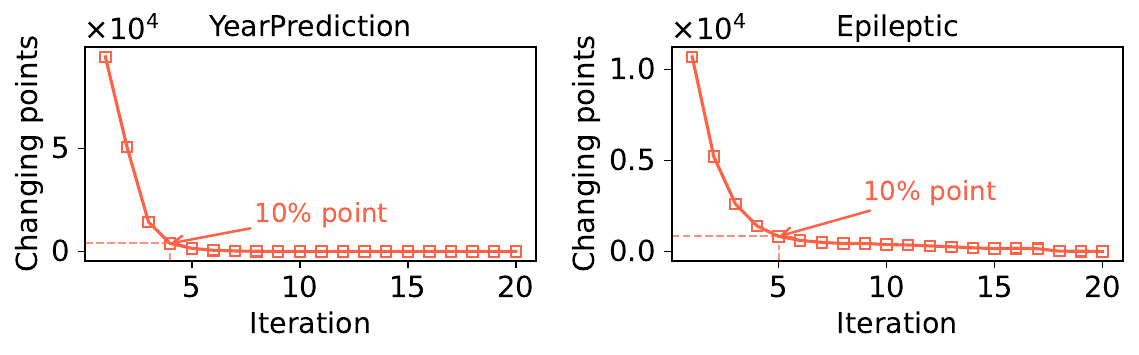}
  \vspace{-2.5em}
  \caption{
  In \CDKM, the number of points with changed assignments drops sharply in the first few iterations, indicating that most points quickly stabilize and allow many later assignment evaluations to be pruned.}
\label{fig:pruning_motivation}
  \vspace{-1em}
\end{figure}

\CDKM~\cite{nie2022CDKM} naturally avoids the above issue by optimizing sample assignments directly, without explicitly computing centroids. 
It reformulates the $k$-means objective as trace minimization and removes explicit centroid variables, making it a useful basis for VFL clustering.
However, the computational cost of \CDKM still grows rapidly with the number of samples, the feature dimension, and clusters.
As shown in Fig.~\ref{fig:pruning_motivation}, the number of points whose assignments change in each iteration drops sharply after only a few iterations, suggesting that many candidate evaluations are redundant.
Existing exact pruning techniques~\cite{Elkan03Using, Hamerly10Making, ding15Yinyang,wang20Efficiency,zhang2024speeding} are designed around explicit centroids and point-to-center bounds, making them difficult to extend to \CDKM, where cluster statistics change after each move.

To close this gap, we develop \TriCD, an exact triangle-inequality-based pruning strategy for the \CDKM update.
For each row update, \TriCD first evaluates the current cluster exactly and then uses cluster-sum vectors and their accumulated drifts to bound candidate inner products.
Candidates whose bounds cannot improve the incumbent are skipped, so pruning preserves the original \CDKM assignment rule.
Unlike centroid-based pruning, \TriCD builds on-demand bounds from cluster-sum statistics, avoiding persistent point-to-centroid bound maintenance and incurring negligible extra space.

Although \TriCD reduces local computation, communication rounds remain a key concern when extending fast $k$-means to VFL.
Existing federated $k$-means frameworks typically either communicate little while incurring costly server-side matching, or avoid such matching at the cost of many communication rounds.
\VFLC~\cite{Ding16KMeans} and \DPFMPS~\cite{li2022differentially} reduce communication by using one-shot summaries, but their server-side matching requires Cartesian-product enumeration over client-side summaries.
\FFFKM~\cite{Zhu23F3KM} adopts block-coordinate optimization, but its communication rounds grow linearly with the dataset size.
These methods expose a key trade-off in VFL clustering: reducing communication rounds may leave substantial cross-client computation, while avoiding such computation often requires more communication rounds.
Hence, an efficient VFL $k$-means algorithm should balance local computation and communication rounds.

To address these challenges, we further develop \FedTriCD, an efficient VFL $k$-means algorithm that avoids explicitly materializing centroids and leverages triangle-inequality-based pruning to reduce redundant computations.
\FedTriCD decomposes the objective into client-local subproblems, so each client optimizes with only local features, applies local bound-based pruning, and uploads objective-value surrogates rather than raw data or centers.
Based on this formulation, \FedTriCD reduces communication rounds by allowing multiple local epochs between server communications and stabilizes optimization with a proximal regularization term that limits local-to-global drift.
Our goal is to improve local and communication rounds while maintaining competitive clustering quality.
Overall, our contributions are as follows:
\begin{itemize}
    \item 
    We propose \TriCD, a triangle-inequality-based exact pruning strategy for \CDKM that accelerates coordinate-wise cluster reassignment by safely eliminating unnecessary objective evaluations (Section~\ref{sec:tricd}).

    \item  We propose \FedTriCD, a VFL clustering algorithm that reduces communication rounds by allowing multiple pruned local updates, while using a proximal regularization term to stabilize optimization and limit local-to-global drift (Section~\ref{sec:fedtricd}). 
    
    \item 
    We provide theoretical guarantees for the proposed methods, proving the exactness and finite convergence of \TriCD and establishing a bounded-error server-side update guarantee for \FedTriCD (Section~\ref{sec:convergence}).

    \item Extensive experiments on real-world datasets show that \TriCD achieves up to an $50.07\times$ speedup over \Lloyd's algorithm, while \FedTriCD achieves a better trade-off among computational efficiency, communication rounds, and clustering quality than state-of-the-art VFL clustering methods (Section~\ref{sec:experiments}).

\end{itemize}

\section{Related Work}
In this section, we review related work in two domains: federated clustering and fast clustering. 

\noindent\textbf{Federated clustering.}
In recent years, vertical federated clustering has attracted increasing attention~\cite{Ding16KMeans, Dennis21Heterogeneity,li2022differentially, Zhu23F3KM, ji2025federated}.
Ding et al.~\cite{Ding16KMeans} proposed a variant of $k$-means in which each client computes local centers based on its own feature subset, and the server aggregates them. 
Huang et al.~\cite{Huang2022Coresets} introduced a general framework for constructing coresets in VFL and applied it to $k$-means. 
Li et al.~\cite{li2022differentially} adopted differentially private $k$-means in VFL, where the server builds a weighted grid to summarize the global dataset and provide privacy guarantees for the learned centers.
Zhu et al.~\cite{Zhu23F3KM} proposed a fair federated $k$-means method and used ADMM~\cite{Stephen2011Distributed} to solve the subproblems, improving both efficiency and fairness. 
Ji et al.~\cite{ji2025federated} incorporated balance constraints through a penalty formulation, but the performance can be sensitive to the penalty parameter. 
Despite recent progress, federated clustering methods still require many communication rounds or incur high computational complexity.

\noindent\textbf{Fast clustering.}
In this work, we focus on accelerating $k$-means by reducing redundant distance computations~\cite{shu2022Ball,wang2020efficiency}. 
Among existing techniques, bound-based pruning is one of the most influential approaches, as it avoids unnecessary evaluations of point-center distances while preserving exact assignment results.
\Elkan~\cite{Elkan03Using} uses triangle-inequality bounds between data points and candidate centroids to safely skip redundant distance computations. 
To reduce the space and maintenance overhead of such per-point multi-centroid bounds, Hamerly~\cite{Hamerly10Making} proposes a lightweight variant that maintains only one lower bound per point while retaining effective pruning. 
\Yinyang~\cite{ding15Yinyang} groups centroids and derives group-wise bounds, while Newling et al.~\cite{newling2016fast} and tree-based methods~\cite{shu2022Ball} accelerate assignment by pruning large sets of candidate centroids in a batch.
However, the effectiveness of such structured or spatial pruning may deteriorate in high-dimensional settings, where geometric separation becomes weaker.
\LBKC~\cite{zhang2024speeding} constructs inexpensive lower bounds of Euclidean distances to prune redundant sample-to-centroid evaluations while preserving exact clustering results.
Existing fast $k$-means methods are based on the Lloyd heuristic.
Because these methods accelerate Lloyd-style assignments, they inherit Lloyd's sensitivity to initialization and its tendency to converge to local optima, and often incur extra bound-maintenance overhead across iterations.

\noindent\textbf{Remarks.}
Existing federated $k$-means methods often incur high computational overhead or require many communication rounds.
Moreover, the requirements of efficient federated optimization further constrain the design space, making existing pruning techniques difficult to apply directly.
These factors make it challenging to achieve both low computation cost and few communication rounds in VFL clustering.


\begin{table}[ht]
    \centering
    \caption {Summary of notation}
    \vspace{-1em}
\label{tab:notations}
\footnotesize
\setlength\tabcolsep{4pt}
\renewcommand{\arraystretch}{1}
\resizebox{0.95\columnwidth}{!}{%
    \begin{tabular}{ll}
    \toprule
    \textbf{Notation} & \textbf{Description} \\
    \midrule
    $\bX\in\mathbb{R}^{d\times n}$ & The dataset \\
    $\mathcal{S}$ & The set of clusters \\
    $\bC\in\mathbb{R}^{d\times k}$ & The center matrix \\
    $\bF\in\{0,1\}^{n\times k}$ & The indicator matrix \\
    $\psi$ & The \CDKM row-update objective \\
    $\phi$ & The client-local row-update objective \\
    $\underline{\phi},\widehat{\phi}$ & The pruning bound and surrogate value \\
    \bottomrule
    \end{tabular}
}
    \vspace{-1 em}
\end{table}

\section{Preliminaries} 
\label{sec:pre}
This section reviews the $k$-means formulation of Nie et al.~\cite{nie2022CDKM}, referred to as \CDKM, and the triangle inequality and the basic VFL setting. 
Table~\ref{tab:notations} summarizes the notations used in this paper.

\subsection{\texorpdfstring{Description of \CDKM}{k-means without Centers}}
\label{sec:cdkm_prelim}
Let $\bX \in \mathbb{R}^{d \times n}$ contain $n$ samples, with column $\bx_i$ denoting the $i$-th data point.
The standard $k$-means objective partitions the samples into $k$ clusters $\mathcal{S}=\{\bS_1,\ldots,\bS_k\}$ by minimizing the sum of squared errors:
\begin{align}\label{eq:SSE1}
\min_{\mathcal{S},\bC}\sum_{j=1}^{k} \sum_{\bx_{i} \in \bS_{j}} ||\bx_{i} - \bc_{j}||_2^{2},
\end{align}
where $\bc_j$ is the centroid of cluster $\bS_j$.
Let $\bF \in \mathbb{R}^{n \times k}$ be the one-hot indicator matrix, where $f_{ij}=1$ if $\bx_i\in\bS_j$ and $f_{ij}=0$ otherwise.
Then Eq.~\eqref{eq:SSE1} can be written as:
\begin{align}\label{eq:SSE2}
\min _{\bF  \in \operatorname{\textit{Ind}}, \bC} \! \sum_{j=1}^{k} \sum_{i=1}^{n}\|\bx_{i}-\bc_{j}\|_{2}^{2} f_{ij}\!\!=\!\!\min_{\bF \in \operatorname{\textit{Ind}}, \bC}\!\|\bX-\bC \bF^{\top}\|_{F}^{2},
\end{align}
where $\operatorname{\textit{Ind}}$ denotes the set of valid one-hot indicator matrices.
For fixed $\bF$, the optimal center matrix is
\begin{align}\label{eq:sloutionC}
\bC=\bX\bF (\bF^{\top} \bF)^{-1}.
\end{align}
Substituting Eq.~\eqref{eq:sloutionC} into Eq.~\eqref{eq:SSE2} and discarding the constant term yields the objective value used by \CDKM.
Since $\bF^\top\bF$ is diagonal for one-hot assignments, this objective can be written as:
\begin{align}\label{eq:obj_cdkm} 
\min _{  \bF  \in \operatorname{\textit{Ind}}} - \sum_{j=1}^{k} \frac{\bff_{j}^{\top} \bX^{\top} \bX \bff_{j}}{\bff_{j}^{\top} \bff_{j}}.
\end{align}

\CDKM minimizes Eq.~\eqref{eq:obj_cdkm} by coordinate descent over rows of $\bF$~\cite{Stephen2015Coordinate}.
For sample $i$, let $\bF_i^{(m)}$ be the assignment obtained by setting row $i$ to cluster $m$ while keeping all other rows fixed.
The corresponding one-row subproblem is
\begin{align}\label{eq:subobj}
\min _{m \in [k]}\psi(\bF_{i}^{(m)})=\min _{m \in [k]}-\sum_{j=1}^{k} \frac{(\bff_{j}^{(m)})^{\top}\bX^{\top} \bX \bff_{j}^{(m)}}{(\bff_{j}^{(m)})^{\top} \bff_{j}^{(m)}}. 
\end{align} 
Let $\bF_i^{(0)}$ denote the same assignment except that row $i$ is set to zero.
Since the cluster-independent part cancels, Eq.~\eqref{eq:subobj} is equivalently written as:
\begin{equation}
\begin{aligned} \label{eq:simplesubobj}
&\min _{m \in[k]} \psi(i, m)  
= \min _{m \in[k]} \psi(\bF_{i}^{(m)})-\psi(\bF_{i}^{(0)})  \\
&= \min _{m \in[k]} \frac{(\bff_{m}^{(0)})^{\top} \bX^{\top} \bX \bff_{m}^{(0)}}{(\bff_{m}^{(0)})^{\top} \bff_{m}^{(0)}}-\frac{(\bff_{m}^{(m)})^{\top} \bX^{\top} \bX \bff_{m}^{(m)}}{(\bff_{m}^{(m)})^{\top} \bff_{m}^{(m)}}.
\end{aligned}
\end{equation}

Let $r$ be the current cluster of sample $i$, and let $\bff_m$ be the $m$-th column of $\bF$.
From Eq.~\eqref{eq:simplesubobj}, the objective value has two forms.
If $m=r$, removing sample $i$ from cluster $r$ gives
\begin{equation}
\label{eq:CDKM_obj_case_1}
\begin{aligned}
\hspace{-1.5em}{\psi}(i, m)  \!
= \!-\!\frac{\bff_{m}^{\top}\bX^{\top} \!\bX \bff_{m}}{\bff_{m}^{\top} \bff_m}
\!+\!\frac{\bff_m^{\top} \bX^{\top} \!\bX \bff_m \! \!- \!2 \bx_i^{\top} \bX \bff_m\!\! + \!\bx_i^{\top} \bx_i}{\bff_m^{\top} \bff_{m}  - 1 }.
\end{aligned}
\end{equation}
If $m\neq r$, adding sample $i$ to candidate cluster $m$ gives
\begin{equation}
\label{eq:CDKM_obj_case_2}
\begin{aligned}
\hspace{-1.8em}{\psi}(i, m) 
= \!\frac{\bff_{m}^{\top}\bX^{\top} \!\bX \bff_{m}}{\bff_{m}^{\top} \bff_m}
\! \!- \!\!\frac{\bff_m^{\top} \bX^{\top} \!\bX \bff_m \! \!+ \!2 \bx_i^{\top} \bX \bff_m\!\! + \!\bx_i^{\top} \bx_i}{\bff_m^{\top} \bff_{m}  + 1 }.
\end{aligned}
\end{equation}

\begingroup
\begin{algorithm}[t]
    \caption{\CDKM}
    \label{alg: CDKM}
    \KwIn{Data $\bX$, clusters $k$, initial assignment $\bF$.}
    \KwOut{Indicator matrix $\bF$.}
    Initialize $\bF$, $\bX \bff_m$, $\bff_m^{\top} \bff_m$, $\bff_m^{\top} \bX^{\top} \bX \bff_m$, $(\bx_j)^{\top} \bx_j$, $m \in [k], j \in [n]$\;
    \While{not converged}{
        \For{$i=1,\dots,n$}{
            Compute $\psi(i,m)$ by Eq.~\eqref{eq:CDKM_obj_case_1} and Eq.~\eqref{eq:CDKM_obj_case_2}\;
            Update $\bF$ by Eq.~\eqref{eq:cdkm_update_f}\;
            Update $\bX\bff_m$, $\bff_m^{\top}\bff_m$, and $\bff_m^{\top}\bX^{\top}\bX\bff_m$ using Eq.~\eqref{eq:cdkm_update_stats}\;
        }
    }
    \KwRet{$\bF$}\;
\end{algorithm}
\endgroup

Using \eqref{eq:CDKM_obj_case_1} and \eqref{eq:CDKM_obj_case_2}, \CDKM updates row $i$ by choosing the cluster with the smallest candidate objective value:
\begin{equation}\label{eq:cdkm_update_f}
f_{iq}=
\begin{cases}
1, & q=\arg \min _{m}  \psi(i, m) ; \\ 
0, & \text { otherwise. }
\end{cases}
\end{equation}
where $q$ is the updated cluster index.
After a nontrivial move from $r$ to $q$, the sufficient statistics are updated incrementally as follows, where the quantities on the right-hand side are evaluated before the update:
\begin{equation}
\label{eq:cdkm_update_stats}
\begin{aligned}
&\hspace{1.5em} \bX \bff_r \gets \bX \bff_r - \bx_i, \bX \bff_q \gets \bX \bff_q + \bx_i, \\
&\hspace{1.5em} \bff_r^{\top} \bff_r \gets \bff_r^{\top} \bff_r - 1, \bff_q^{\top} \bff_q \gets \bff_q^{\top} \bff_q + 1, \\
&\bff_r^{\top} \bX^{\top} \bX \bff_r \gets \bff_r^{\top} \bX^{\top} \bX \bff_r - 2 \bx_i^{\top} \bX \bff_r + \bx_i^{\top} \bx_i, \\
&\bff_q^{\top} \bX^{\top} \bX \bff_q \gets \bff_q^{\top} \bX^{\top} \bX \bff_q + 2 \bx_i^{\top} \bX \bff_q + \bx_i^{\top} \bx_i.
\end{aligned}
\end{equation}

\subsection{Triangle Inequality}
The triangle inequality is widely used to accelerate machine learning and similarity search by bounding expensive distance computations~\cite{paparrizos2022fast, schubert2021triangle, wang2011fast,ding2017generalizations, Elkan03Using, Hamerly10Making, ding15Yinyang}.
Its basic form is
\begin{equation}
\begin{aligned}
 l(A,C) \leq l(A,B) + l(B,C),
\label{eq:triangle_inequality}
\end{aligned}
\end{equation}
where $A$, $B$, and $C$ are points in a Euclidean space, and $l$ denotes the Euclidean distance.
It bounds an unknown distance in terms of known distances, enabling safe pruning without directly evaluating every candidate.

\subsection{Vertical Federated Learning Setting}
\label{sec:vfl_prelim}
In VFL with $H$ clients, each client $h$ holds a feature subset $\bX_h \in \mathbb{R}^{d_h \times n}$ for the same $n$ samples and collaborates through a central server to obtain a global clustering assignment $\bF$.
The global objective is
\begin{equation}
    \begin{aligned}
    \min_{\bF} G(\bF) = \sum_{h=1}^{H} \alpha_h g_h(\bF; \bX_h),
    \end{aligned}
    \end{equation}
where $\alpha_h$ is typically chosen as $1/H$ or $d_h/d$, with $d=\sum_{h=1}^{H} d_h$.
The term $g_h(\bF; \bX_h)$ denotes the local SSE contribution of client $h$.


\section{Algorithm Design}

In this section, we aim to address the following challenges:
\begin{itemize}

  \item How can we exactly prune redundant inner-product evaluations to accelerate clustering without altering the exact coordinate-update rule?

  \item How can we decompose the $k$-means objective into client-local subproblems that can be solved independently while reducing communication rounds?
  
\end{itemize}

To address these challenges, we propose the following strategies:
1) We develop \TriCD, an exact triangle-inequality-based pruning algorithm that eliminates redundant evaluations through on-demand bounds without changing the original update rule (Section~\ref{sec:tricd}).
2) We decompose the fast $k$-means problem into client-local subproblems (Section~\ref{sec:fedtricd}).
3) We solve these subproblems with periodic local updates to reduce communication rounds, while adding proximal regularization to control client drift (Section~\ref{sec:fedtricd}).

\subsection{Proposed \TriCD} 
\label{sec:tricd}
Next, we present \TriCD, an exact pruning $k$-means algorithm based on the triangle inequality.

\begingroup
\begin{algorithm}[t]
\caption{\TriCD}
\label{alg:tricd}
\KwIn{Data $\bX$, clusters $k$, initial $\bF$.}
\KwOut{Indicator matrix $\bF$.}
Init $\bb_m$, $a_m$, $\|\bb_m\|_2^2$, $\|\bx_i\|_2^2$, $m \in [k], i \in [n]$\;
\While{not converged}{
Compute $D_{rm}$ for all $r,m\in[k]$\;
Set $\Delta\bb_m\leftarrow 0$ and $\delta_m\leftarrow 0$, $m \in [k]$\;
\For{$i=1,\dots,n$}{
Set $\psi^\star\leftarrow\psi(i,r)$ by Eq.~\eqref{eq:tricd_psi}\;
\For{$m\in[k]\setminus\{r\}$}{
Set $ub_{im}$ by Eq.~\eqref{eq:tri_interval_new}\;
Compute $\underline{\psi}(i,m)$ by Eq.~\eqref{eq:psi_lower_bound}\;
\If{$\underline{\psi}(i,m)<\psi^\star$}{
Compute $\psi(i,m)$ by Eq.~\eqref{eq:tricd_psi}\;
}
}
Update $\bF$, $\bb_m$, $a_m$, $\|\bb_m\|_2^2$, $\delta_m$  by Eq.~\eqref{eq:TriCD_Update_Stats};
}
}
\KwRet{$\bF$}\;
\end{algorithm}
\endgroup

\myparagraph{Triangle-Inequality-based Pruning for \CDKM}
\TriCD accelerates the one-row update in Eq.~\eqref{eq:simplesubobj} by evaluating the current cluster exactly and using triangle-inequality-based lower bounds to skip candidates that cannot improve the incumbent.
For each cluster $m$, let $a_m = \bff_m^\top \bff_m$ be its size and $\bb_m = \bX \bff_m$ be its sum vector.
We also maintain $\|\bb_m\|_2^2$ incrementally.
Let $s_{im}= \bx_i^\top \bb_m$, and $r$ be the current cluster of $\bx_i$.
The one-row score is
\begin{equation}
\psi(i,m)=
\begin{cases}
\displaystyle
\frac{\|\bb_m\|_2^2 - 2 s_{im} + \|\bx_i\|_2^2}{a_m-1} - \frac{\|\bb_m\|_2^2}{a_m},
& m=r,\\[9pt]
\displaystyle
\frac{\|\bb_m\|_2^2}{a_m} - \frac{\|\bb_m\|_2^2 + 2 s_{im} + \|\bx_i\|_2^2}{a_m+1},
& m\neq r.
\end{cases}
\label{eq:tricd_psi}
\end{equation}
All terms in Eq.~\eqref{eq:tricd_psi} are either precomputed or maintained in $\mathcal{O}(1)$,
except the inner product $s_{im}$.
\TriCD avoids unnecessary exact evaluations of $s_{im}$ by maintaining upper bounds on $s_{im}$ that induce valid lower bounds on candidate scores.

\TriCD maintains a valid upper bound $ub_{im}$ on $s_{im}$ for the current row update.
It satisfies
\begin{equation}
s_{im} \le ub_{im}, \qquad i\in[n],~ m\in[k].
\label{eq:bound_def}
\end{equation}

During each iteration, reassignment changes the source and destination cluster sums.
We therefore record each cluster sum at the beginning of the iteration and track its drift:
\begin{align}
\Delta \bb_r &= \bb_r - \bb_r^{(0)}, \qquad \delta_r = \|\Delta \bb_r\|_2, \\
\Delta \bb_m &= \bb_m - \bb_m^{(0)}, \qquad \!\!\!\delta_m = \|\Delta \bb_m\|_2,
\label{eq:drift_def}
\end{align}

For point $\bx_i$ with current cluster $r$, we compute $s_{ir}$ exactly and set $ub_{ir}=s_{ir}$. 
For any candidate cluster $m\neq r$,
\begin{equation}
s_{im}-s_{ir}=\bx_i^\top(\bb_m-\bb_r),
\end{equation}
and the Cauchy-Schwarz inequality gives
\begin{equation}
|s_{im}-s_{ir}| \le \|\bx_i\|_2 \cdot \|\bb_m-\bb_r\|_2.
\label{eq:tri_inner_prod}
\end{equation}
We compute the reference distance between cluster-sum vectors, $D_{rm}=\|\bb_r^{(0)}-\bb_m^{(0)}\|_2$ once at the start of the iteration.

\begin{figure}[t]
    \centering
    \includegraphics[width=0.85\linewidth]{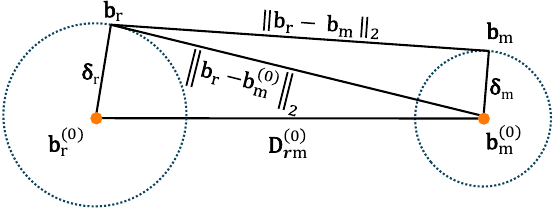}
      \vspace{-1.0em}
    \caption{Triangle-inequality bound for $\|\bb_r-\bb_m\|_2$.}
      \vspace{-1.0em}
    \label{fig:tricd_drift_triangle}
\end{figure}

As illustrated in Fig.~\ref{fig:tricd_drift_triangle}, the current inter-cluster distance is bounded by the initial distance and the two drifts:
\begin{align}
\|\bb_r-\bb_m\|_2
&\le \|\bb_r-\bb_m^{(0)}\|_2 + \|\bb_m^{(0)}-\bb_m\|_2 \nonumber\\
&\le \|\bb_r-\bb_r^{(0)}\|_2 + \|\bb_r^{(0)}-\bb_m^{(0)}\|_2 + \|\bb_m^{(0)}-\bb_m\|_2 \nonumber\\
&= D_{rm} + \delta_r + \delta_m .
\label{eq:safe_dist_new}
\end{align}
Combining Eq.~\eqref{eq:tri_inner_prod} and Eq.~\eqref{eq:safe_dist_new} gives
\begin{equation}
s_{im} \le
s_{ir}+\|\bx_i\|_2 (D_{rm}+\delta_r+\delta_m),
\label{eq:tri_interval_new}
\end{equation}
and we set $ub_{im}$ to the right-hand side of Eq.~\eqref{eq:tri_interval_new}.

For $m\neq r$ in Eq.~\eqref{eq:tricd_psi}, $\psi(i,m)$ is monotonically decreasing in $s_{im}$ when $\|\bb_m\|_2^2,a_m,$ and $\|\bx_i\|_2^2$ are fixed.
Hence substituting the upper bound $ub_{im}$ yields a valid lower bound:
\begin{equation}
 \! \! \!\underline{\psi}(i,m)\!=\!\frac{\|\bb_m\|_2^2}{a_m}
 \!-\! \frac{\|\bb_m\|_2^2 + 2\,ub_{im} + \|\bx_i\|_2^2}{a_m+1} \le \psi(i,m).
\label{eq:psi_lower_bound}
\end{equation}
Let $\psi^\star$ denote the incumbent smallest exact score for point $i$.
\TriCD first evaluates $\psi(i,r)$ exactly (if $a_r>1$) to initialize $\psi^\star$.
For each $m\neq r$, it prunes $m$ when $\underline{\psi}(i,m)\ge \psi^\star$ and evaluates $s_{im}$ exactly only when $\underline{\psi}(i,m) < \psi^\star$.

Thus \TriCD returns the same minimizer as Eq.~\eqref{eq:tricd_psi} while avoiding unnecessary exact inner products.
Accordingly, the update function for the $i$-th row of $\bF$ is
\begin{equation}\label{eq:tricd_update_f}
f_{iq}=
\begin{cases}
1, & q=\arg \min_{m} \psi(i,m),\\
0, & \text{otherwise,}
\end{cases}
\end{equation}
where $f_{iq}$ is the $(i,q)$ element of $\bF$.
When $r=q$, $\bF$ remains unchanged.
For a move from $r$ to $q$, \TriCD uses the same incremental statistics update as \CDKM, with all quantities on the right evaluated before moving sample $i$:
\begin{equation}
\begin{gathered}
a_r \gets a_r - 1; a_q \gets a_q + 1;\\
\bb_r \gets \bb_r - \bx_i; \bb_q \gets \bb_q + \bx_i;\\
\Delta\bb_r \gets \Delta\bb_r - \bx_i; \Delta\bb_q \gets \Delta\bb_q + \bx_i;\\
\delta_r \gets \|\Delta\bb_r\|_2; \delta_q \gets \|\Delta\bb_q\|_2;\\
\|\bb_r\|_2^2 \gets \|\bb_r\|_2^2 - 2\;\bx_i^\top \bb_r + \|\bx_i\|_2^2;\\
\|\bb_q\|_2^2 \gets \|\bb_q\|_2^2 + 2\;\bx_i^\top \bb_q + \|\bx_i\|_2^2.
\end{gathered}
\label{eq:TriCD_Update_Stats}
\end{equation}
Algorithm~\ref{alg:tricd} provides a detailed description of \TriCD.
Its exactness proof is provided in Section~\ref{sec:tricd_convergence}.

\noindent\textbf{Remarks.}
\TriCD keeps the original \CDKM update while reducing exact evaluations of $s_{im}$.
To avoid empty clusters, a point is not reassigned when its current cluster has size $a_r=1$.

\begin{figure}[t]
  \centering
  \includegraphics[width=\linewidth]{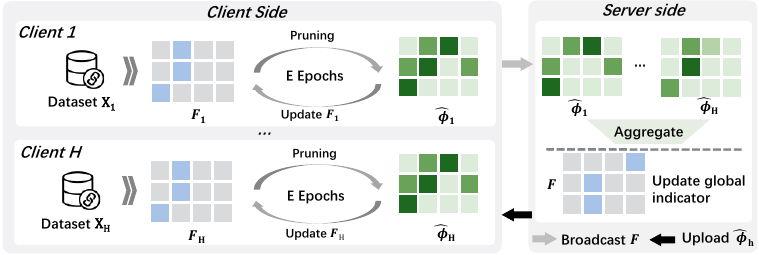}
  \vspace{-1.8em}
  \caption{One communication round of \FedTriCD.}
  \vspace{-1em}
  \label{fig:fedtricd_framework}
\end{figure}

\subsection{Proposed \FedTriCD}
\label{sec:fedtricd}

We introduce \FedTriCD, an efficient VFL $k$-means algorithm that avoids explicitly materializing centroids and leverages triangle-inequality-based pruning to reduce redundant computations.
\FedTriCD decomposes the $k$-means update into client-local subproblems and solves them with pruning.
To reduce communication rounds, each client performs $E$ local epochs before communicating with the server, while a proximal term limits the drift from the global assignment.
Fig.~\ref{fig:fedtricd_framework} illustrates one communication round of \FedTriCD.

\myparagraph{VFL decomposition and local objective}
We consider a VFL system with $H$ clients and a server.
We write $\{\bX_h\}_{h=1}^H$ for the feature-wise partition of $\bX$ over $H$ clients. 
Each client shares the same $n$ aligned samples and holds $\bX_h\in\mathbb{R}^{d_h\times n}$, with $\sum_{h=1}^H d_h=d$.
Here, $\bF$ is the global indicator matrix, and $\bF_h$ is the local indicator matrix on client $h$.
Let $\bff_m^{(h)}$ be the $m$-th column of $\bF_h$ and set $a_m^{(h)}=(\bff_m^{(h)})^\top\bff_m^{(h)}$, $\bb_m^{(h)}=\bX_h\bff_m^{(h)}$, and
$s_{im}^{(h)}=(\bx_i^{(h)})^\top\bb_m^{(h)}$.
Then the statistics in Eq.~\eqref{eq:tricd_psi} decompose as:
\begin{align}
&s_{im}=\bx_i^\top\bb_m
= \sum_{h=1}^H (\bx_i^{(h)})^\top \bb_m^{(h)}
= \sum_{h=1}^H s_{im}^{(h)}.
\label{eq:fedtricd_dist_decompose}
\end{align}
Let $r$ be the current local cluster of sample $i$.
As in \TriCD, the nonempty-cluster rule is applied locally: if $a_r^{(h)}=1$, client $h$ keeps sample $i$ in cluster $r$.
The local score for row $i$ and candidate cluster $m$ is
\begin{equation}
\psi(i,m,h)\!=\!
\begin{cases}
\displaystyle
\frac{\|\bb_m^{(h)}\|_2^2 \!-\! 2 s_{im}^{(h)}\! +\! \|\bx_i^{(h)}\|_2^2}{a_m^{(h)}-1} \!- \!\frac{\|\bb_m^{(h)}\|_2^2}{a_m^{(h)}},
& m=r,\\[9pt]
\displaystyle
\frac{\|\bb_m^{(h)}\|_2^2}{a_m^{(h)}} \!-\! \frac{\|\bb_m^{(h)}\|_2^2 \!+\! 2 s_{im}^{(h)} \!+\! \|\bx_i^{(h)}\|_2^2}{a_m^{(h)}+1},
& m\neq r.
\end{cases}
\label{eq:fedtricd_additive_psi}
\end{equation}

\myparagraph{Proximal local subproblem}
Let $m_i^{\mathrm{g}}$ be the  cluster of sample $i$ in $\bF$.
For a row update, the proximal term adds $\rho\,\mathbb{I}\{m\neq m_i^{\mathrm{g}}\}$.
Substituting it into Eq.~\eqref{eq:fedtricd_additive_psi}, we have
\begin{equation}
\phi(i,m,h)
=
\psi(i,m,h)
+\rho\,\mathbb{I}\{m\neq m_i^{\mathrm{g}}\},
\label{eq:fedtricd_prox}
\end{equation}
where $\rho>0$ controls how strongly $\bF_h$ is tied to $\bF$.

\myparagraph{Client-local pruning}
The only expensive term in Eq.~\eqref{eq:fedtricd_prox} is the local inner product $s_{im}^{(h)}$.
At the beginning of each local epoch, client $h$ records local cluster sums, computes $D_{rm}^{(h)}=\|\bb_r^{(h,0)}-\bb_m^{(h,0)}\|_2$, and tracks drifts $\delta_m^{(h)}$.
After evaluating the current cluster exactly, client $h$ bounds each candidate $m\neq r$ by
$ub_{im}^{(h)}=s_{ir}^{(h)}+\|\bx_i^{(h)}\|_2(D_{rm}^{(h)}+\delta_r^{(h)}+\delta_m^{(h)})$.
This upper bound and the known proximal penalty yield
\begin{equation}
\begin{aligned}
\underline{\phi}(i,m,h)
=&
\frac{\|\bb_m^{(h)}\|_2^2}{a_m^{(h)}}
-\frac{\|\bb_m^{(h)}\|_2^2 + 2\,ub_{im}^{(h)} + \|\bx_i^{(h)}\|_2^2}{a_m^{(h)}+1}\\
&+\rho\,\mathbb{I}\{m\neq m_i^{\mathrm{g}}\}
\le \phi(i,m,h).
\end{aligned}
\label{eq:fedtricd_lower_prox}
\end{equation}
The uploaded value is
\begin{equation}
\widehat{\phi}(i,m,h)=
\begin{cases}
\underline{\phi}(i,m,h), & \text{if }m\text{ is pruned},\\
\phi(i,m,h), & \text{otherwise}.
\end{cases}
\label{eq:fedtricd_surrogate_prox}
\end{equation}

\myparagraph{Client update $\bF_h$}
Client $h$ updates row $i$ of $\bF_h$ by
\begin{equation}\label{eq:fedtricd_local_update_f}
f_{iq}^h =
\begin{cases}
1, & q = \arg \min_{m} \, \phi(i, m, h), \\
0, & \text{otherwise}.
\end{cases}
\end{equation}
For a nontrivial local move from $r$ to $q$, client $h$ updates its local statistics as below, with all quantities on the right evaluated before moving sample $i$:
\begin{equation}
\begin{gathered}
a_r^{(h)} \gets a_r^{(h)} - 1;\quad a_q^{(h)} \gets a_q^{(h)} + 1; \\
\bb_r^{(h)} \gets \bb_r^{(h)} - \bx_i^{(h)}; \bb_q^{(h)} \gets \bb_q^{(h)} + \bx_i^{(h)};\\
\Delta\bb_r^{(h)} \gets \Delta\bb_r^{(h)} - \bx_i^{(h)}; \Delta\bb_q^{(h)} \gets \Delta\bb_q^{(h)} + \bx_i^{(h)};\\
\delta_r^{(h)} \gets \|\Delta\bb_r^{(h)}\|_2; \delta_q^{(h)} \gets \|\Delta\bb_q^{(h)}\|_2;\\
\|\bb_r^{(h)}\|_2^2 \gets \|\bb_r^{(h)}\|_2^2 - 2\;\bx_i^{(h)\top}\bb_r^{(h)} + \|\bx_i^{(h)}\|_2^2;\\
\|\bb_q^{(h)}\|_2^2 \gets \|\bb_q^{(h)}\|_2^2 + 2\;\bx_i^{(h)\top}\bb_q^{(h)} + \|\bx_i^{(h)}\|_2^2.
\end{gathered}
\label{eq:fedtricd_update_stats}
\end{equation}

\noindent Algorithm~\ref{alg:FedTriCD} summarizes the full procedure.
\begingroup
\begin{algorithm}[t]
\caption{\FedTriCD}
\label{alg:FedTriCD}
\KwIn{Clients $\{\bX_h\}_{h=1}^H$, clusters $k$, local steps $E$, proximal coefficient $\rho$.}
\KwOut{Indicator matrix  $\bF$.}
\textcolor{gray}{\underline{\textit{Server broadcast:}}} Initialize and send $\bF$ to all clients\nllabel{line:fedtricd_server_init}\;
\textcolor{gray}{\underline{\textit{Client init:}}} Set $\bF_h\leftarrow\bF$ and initialize $a_m^{(h)}$, $\bb_m^{(h)}$, $\|\bb_m^{(h)}\|_2^2$, $\|\bx_i^{(h)}\|_2^2$, $m \in [k]$, $i \in [n]$\nllabel{line:fedtricd_client_init}\;
\While{not converged}{
\For{all $H$ clients in parallel}{
\For{$e=1,\dots,E$}{
\textcolor{gray}{\underline{\textit{Client pre-computation:}}} Compute $D_{rm}^{(h)}$ and initialize $\Delta\bb_m^{(h)}$ and $\delta_m^{(h)}$ for all $r,m\in[k]$\nllabel{line:fedtricd_client_init}\;
\textcolor{gray}{\underline{\textit{Client computation:}}}
\For{$i=1,\dots,n$}{
Set $\phi_h^\star$ and $\widehat{\phi}(i,r,h)$ by Eq.~\eqref{eq:fedtricd_prox}\;
\For{$m\in[k]\setminus\{r\}$}{
Set $ub_{im}^{(h)}$ by Eq.~\eqref{eq:tri_interval_new}\;
Compute $\underline{\phi}(i,m,h)$ by Eq.~\eqref{eq:fedtricd_lower_prox}\;
\If{$\underline{\phi}(i,m,h)<\phi_h^\star$}{
Compute $\phi(i,m,h)$ by Eq.~\eqref{eq:fedtricd_prox}\;
Set $\widehat{\phi}(i,m,h)\leftarrow \phi(i,m,h)$\;
\If{$\widehat{\phi}(i,m,h)<\phi_h^\star$}{update $\phi_h^\star$\;}
}
\Else(\tcp*[f]{Pruning}){
    Form $\widehat{\phi}(i,m,h)$ by Eq.~\eqref{eq:fedtricd_surrogate_prox}\;
}
}
\textcolor{gray}{\underline{\textit{Client update:}}} Update $\bF_h$ by Eq.~\eqref{eq:fedtricd_local_update_f}\;
\textcolor{gray}{\underline{\textit{Client update:}}} Update client-local statistics by Eq.~\eqref{eq:fedtricd_update_stats}\;
}
}
\textcolor{gray}{\underline{\textit{Client upload:}}} Send $\widehat{\phi}$ to the server\nllabel{line:fedtricd_upload}\;
}
\textcolor{gray}{\underline{\textit{Server aggregation:}}} Aggregate $\widehat{\phi}$\nllabel{line:fedtricd_aggregate}\;
\textcolor{gray}{\underline{\textit{Server update:}}} Sequentially update $\bF$ row by row via Eq.~\eqref{eq:fedtricd_update_f}\nllabel{line:fedtricd_server_update}\;
\For{All H clients in parallel}{
\textcolor{gray}{\underline{\textit{Client download:}}} Receive $\bF$ and set $\bF_h$\nllabel{line:fedtricd_download}\;
\textcolor{gray}{\underline{\textit{Client update:}}} Refresh client-local statistics by Eq.~\eqref{eq:fedtricd_update_stats}\nllabel{line:fedtricd_local_update}\;
}
}
\KwRet{$\bF$}\;
\end{algorithm}
\endgroup

\myparagraph{Server update $\bF$}
After completing $E$ local epochs, clients upload $\widehat{\phi}$ to the server.
The server sequentially scans samples $i=1,\dots,n$ and updates one row of $\bF$ at a time.
For sample $i$, we present the update formulation as follows:
\begin{equation}\label{eq:fedtricd_update_f}
f_{iq}=
\begin{cases}
1, & q=\arg \min _{m} \sum_{h=1}^{H}{\widehat{\phi}(i, m, h)}; \\ 
0, & \text { otherwise,}
\end{cases}
\end{equation}
where ties keep the current assignment.
The server applies the same nonempty-cluster rule with the global assignment.
The server broadcasts the updated $\bF$; each client sets $\bF_h \leftarrow \bF$ and refreshes local statistics row by row using Eq.~\eqref{eq:fedtricd_update_stats}.
Since uploads may include surrogate values and local assignments may drift during the $E$ epochs, the server update is approximate; Section~\ref{sec:convergence} analyzes this error.

\noindent\textbf{Remarks.}
\FedTriCD performs local pruning at each client and employs proximal surrogate aggregation to reduce server-client synchronization rounds while controlling local drift.
At the server, $\bF$ is updated row by row, and moves that would leave any cluster empty are excluded.

\subsection{Complexity Analysis} 
We analyze the communication, computational, and space complexity of \FedTriCD.

\myparagraph{Computational complexity of \FedTriCD}
\label{sec:complexity}
Given $\bX_h\in\mathbb{R}^{d_h\times n}$ and $\bF$, we count per-client multiplications and divisions in Algorithm~\ref{alg:FedTriCD}, ignoring additions.
The initialization computes $\bX_h\bff_m$, $\bff_m^\top\bff_m$, $\bff_m^\top\bX_h^\top\bX_h\bff_m$, and $(\bx_i^{(h)})^\top\bx_i^{(h)}$, requiring $\mathcal{O}(nd_hk+nk+d_hk+nd_h)$ multiplications.

Let $T$ denote the number of communication rounds, and let each round run $E$ local epochs.
In each local epoch, the repeated computation consists of four parts.
First, pre-computing the local reference distances $D_{rm}^{(h)}$ requires $\mathcal{O}(k^2d_h)$ operations.
Second, constructing the current upper bounds and forming lower-bound scores over all candidates require $\mathcal{O}(nk)$ operations.
Third, let $\kappa^{(h)} \in [1,k]$ denote the average number of candidate clusters evaluated exactly on client $h$ after pruning.
The dominant cost of exact local candidate evaluation is $\mathcal{O}(nd_h\kappa^{(h)})$ per local epoch.
Fourth, updating local statistics after row reassignments requires $\mathcal{O}(nd_h)$ operations.
After the $E$ local epochs, server-side aggregation across $H$ clients requires $\mathcal{O}(nHk)$ additions per round.

Therefore, over $T$ communication rounds, the total per-client computational complexity is $\mathcal{O}(nd_hk+nk+d_hk+nd_h)+\mathcal{O}\big((k^2d_h+nk+nd_h\kappa^{(h)}+nd_h)TE\big)$.
When the exact evaluation term dominates the lower-order terms, the per-client cost simplifies to $\mathcal{O}(nd_h\kappa^{(h)}TE)$.
In the worst case where $\kappa^{(h)} = k$, this becomes $\mathcal{O}(n d_h k\,T E)$.

\myparagraph{Communication round complexity}
We focus on the number of communication rounds, which is a common bottleneck in cross-silo VFL deployments.
With a local period $E\ge1$, \FedTriCD communicates once per global round while performing multiple local updates.
Thus, over $T$ global rounds, the communication round complexity is $\mathcal{O}(T)$.
Equivalently, for a fixed total number of local epochs, increasing $E$ reduces the number of server communication rounds by a factor of $E$.

\myparagraph{Space complexity}
We analyze the additional memory footprint of \FedTriCD.
Each client stores local cluster sums and sizes, squared norms, drift statistics, the local distance matrix $D^{(h)}$, uploaded surrogate values, and label-vector assignments.
These components require $\mathcal{O}(d_hk)$, $\mathcal{O}(k+n)$, $\mathcal{O}(d_hk+k)$, $\mathcal{O}(k^2)$, $\mathcal{O}(nk)$, and $\mathcal{O}(n)$ space, respectively.
Therefore, the additional space complexity of \FedTriCD on each client is
$\mathcal{O}(nk + d_hk + k^2 + n)$.
Here, the $\mathcal{O}(nk)$ term comes from the surrogate values uploaded to the server rather than from persistent upper-bound storage; the current upper bounds are constructed on the fly during candidate evaluation.

\section{Convergence and Stability Analysis}
\label{sec:convergence}

This section analyzes the optimization guarantees of \TriCD and \FedTriCD.
We prove the finite convergence of \TriCD and provide a bounded-error stability guarantee for the server update in \FedTriCD.

\subsection{Objective and One-Row Moves}
\label{sec:objective_moves}

Let $\mathcal{F}$ be the set of one-hot indicator matrices with no empty clusters, i.e., $\bF\mathbf{1}_k=\mathbf{1}_n$ and $\bF^\top\bF \succ 0$.
For $\bF\in\mathcal{F}$, let $r_i$ denote the current cluster of sample $i$ and let $a_m=\bff_m^\top\bff_m$.
The feasible candidate set for updating row $i$ is
\begin{equation}
\mathcal{M}_i(\bF)=\{m\in[k]: m=r_i \text{ or } a_{r_i}>1\}.
\label{eq:app_feasible_candidates}
\end{equation}
For $m\in\mathcal{M}_i(\bF)$, we use the candidate objective value $\psi(i,m)$ defined in Eq.~\eqref{eq:simplesubobj}, evaluated under the current assignment $\bF$.
When $a_{r_i}=1$, the feasible set reduces to $\mathcal{M}_i(\bF)=\{r_i\}$; no objective comparison is needed, and the update keeps the current row unchanged.
When $a_{r_i}>1$, $\psi(i,m)$ subtracts the same zero-row baseline $\psi(\bF_i^{(0)})$ for all candidates, for any $m\in\mathcal{M}_i(\bF)$,
\begin{equation}
\psi(\bF_i^{(m)})-\psi(\bF_i^{(r_i)})
=
\psi(i,m)-\psi(i,r_i).
\label{eq:app_baseline_cancel}
\end{equation}
Thus, selecting the cluster with the smallest $\psi(i,m)$ over $\mathcal{M}_i(\bF)$ is equivalent to selecting the feasible one-row candidate with the smallest objective value.
Throughout the analysis, ties are resolved deterministically by keeping the current cluster whenever it is a minimizer.
Equivalently, candidates are scanned in a fixed order and the incumbent is changed only by a strict improvement.

\begin{lemma}[Finite assignment space and objective values]
\label{lemma:app_finite_range}
The feasible assignment set $\mathcal{F}$ is finite with $|\mathcal{F}|\le k^n$.
Consequently, the objective values in Eq.~\eqref{eq:obj_cdkm} over $\mathcal{F}$ form a finite set.
\end{lemma}

\begin{proof}
By definition, $\mathcal{F}\subset\{0,1\}^{n\times k}$ and each row of any $\bF\in\mathcal{F}$ contains exactly one nonzero entry.
Thus, each row has at most $k$ possible assignment patterns, and the $n$ rows have at most $k^n$ joint patterns.
The nonempty-cluster constraint $\bF^\top\bF\succ0$ only removes some of these assignments, so $|\mathcal{F}|\le k^n$.
Since the objective value in Eq.~\eqref{eq:obj_cdkm} is a real-valued function on a finite domain, its values over $\mathcal{F}$ form a finite set.
\end{proof}

\begin{lemma}[Strict descent of exact row updates]
\label{lemma:app_exact_descent}
For a current assignment $\bF\in\mathcal{F}$ and sample $i$, let
\begin{equation}
q_i=\arg\min_{m\in\mathcal{M}_i(\bF)} \psi(i,m)
\end{equation}
under the deterministic tie-breaking rule above.
If $q_i\neq r_i$, then the corresponding one-row update strictly decreases the objective value in Eq.~\eqref{eq:obj_cdkm}.
\end{lemma}

\begin{proof}
Because the current cluster $r_i$ belongs to $\mathcal{M}_i(\bF)$, it is always a feasible candidate.
If $q_i\neq r_i$, the tie-breaking rule implies
\begin{equation}
\psi(i,q_i)<\psi(i,r_i).
\end{equation}
By the baseline cancellation in Eq.~\eqref{eq:app_baseline_cancel}, this is equivalent to
\begin{equation}
\psi(\bF_i^{(q_i)})<\psi(\bF_i^{(r_i)}),
\end{equation}
so the objective value strictly decreases after assigning the sample $i$ to $q_i$.
\end{proof}

\begin{theorem}[Finite convergence of exact coordinate descent]
\label{theorem:app_exact_cd_convergence}
Starting from any $\bF^{(0)}\in\mathcal{F}$, the exact coordinate-descent updates used by \CDKM converge in finitely many nontrivial row updates to a stationary assignment $\bF^\star$, where
\begin{equation}
\psi(i,m)\ge \psi(i,r_i^\star)
\end{equation}
for every sample $i$ and every $m\in\mathcal{M}_i(\bF^\star)$, with $r_i^\star$ denoting the cluster of sample $i$ in $\bF^\star$.
\end{theorem}

\begin{proof}
By Lemma~\ref{lemma:app_exact_descent}, every nontrivial exact row update strictly decreases the objective value.
By Lemma~\ref{lemma:app_finite_range}, this objective takes only finitely many values over $\mathcal{F}$.
Therefore, infinitely many nontrivial row updates are impossible.
After finitely many such updates, no feasible row move can strictly decrease the objective; otherwise another nontrivial update would occur.
Hence, the algorithm reaches a stationary assignment satisfying the stated condition.
\end{proof}

\subsection{Exact Pruning in \TriCD}
\label{sec:tricd_convergence}

\TriCD changes only the way candidate objective values are evaluated.
It evaluates the current cluster exactly, constructs valid lower bounds for other candidate clusters, and prunes only candidates that cannot beat the current incumbent exact objective value.

\begin{theorem}[Correctness of exact pruning]
\label{theorem:tricd_exact_pruning}
Given the dataset $\bX$, \TriCD returns the same row update as \CDKM under the same processing order, initialization, and deterministic tie-breaking rule.
\end{theorem}

\begin{proof}
For any cluster $m\neq r$, $\underline{\psi}(i,m)$ in Eq.~\eqref{eq:psi_lower_bound} is a valid lower bound of $\psi(i,m)$.
Indeed, for fixed $\|\bb_m\|_2^2$, $a_m$, and $\|\bx_i\|_2^2$, the objective value $\psi(i,m)$ is monotonically decreasing in $s_{im}$.
Since $s_{im}\le ub_{im}$, replacing $s_{im}$ with $ub_{im}$ gives
\begin{equation}
\underline{\psi}(i,m) \le \psi(i,m).
\end{equation}

If the current cluster has size one, both \CDKM and \TriCD keep the row unchanged.
Otherwise, let $\psi^\star$ be the incumbent exact objective value, initialized by evaluating the current cluster $r$.
If a candidate $m\neq r$ is pruned, then $\underline{\psi}(i,m)\ge \psi^\star$.
Since $\underline{\psi}(i,m)$ is a valid lower bound,
\begin{equation}
\psi(i,m)\ge \underline{\psi}(i,m)\ge \psi^\star .
\end{equation}
Thus, the pruned candidate cannot improve the incumbent.

All candidates that are not pruned are evaluated exactly.
Therefore, \TriCD discards only candidates that cannot improve the incumbent's exact objective value, and the returned cluster attains the same minimum value as the exact \CDKM update.
Under the same deterministic tie-breaking rule, \TriCD returns the same result as \CDKM.
\end{proof}

\begin{theorem}[Convergence of \TriCD]
\label{theorem:app_tricd_convergence}
Under the same initialization, processing order, and tie-breaking rule, \TriCD generates the same assignment sequence as \CDKM.
Consequently, \TriCD converges finitely to a stationary assignment.
\end{theorem}

\begin{proof}
Theorem~\ref{theorem:tricd_exact_pruning} shows that \TriCD returns the same row minimizer as \CDKM for every sample update.
Therefore, the two methods generate the same sequence of assignments under the same initialization and processing order.
The finite convergence statement then follows directly from Theorem~\ref{theorem:app_exact_cd_convergence}.
\end{proof}

\subsection{Surrogate Error in \FedTriCD Aggregation}
\label{sec:surrogate_error}

In \FedTriCD, clients upload $\widehat{\phi}$ after local pruning.
For sample $i$ and candidate cluster $m$, define the exact and uploaded proximal aggregate scores as
\begin{equation}
\begin{aligned}
\!\!\Phi_i^{(t)}(m)
\!=\!\sum_{h=1}^H \phi^{(t)}(i,m,h),
\widehat{\Phi}_i^{(t)}(m)
\!=\!\sum_{h=1}^H \widehat{\phi}^{(t)}(i,m,h).
\end{aligned}
\label{eq:app_fedtricd_aggregate_scores}
\end{equation}
The server updates $\bF$ using $\widehat{\Phi}_i^{(t)}(m)$, while $\Phi_i^{(t)}(m)$ is the exact proximal score induced by the same local states.
The difference between them comes from replacing pruned exact scores by lower-bound surrogate values; its magnitude is affected by the local-state drift accumulated during the $E$ local epochs.
Define the server-side surrogate error as
\begin{equation}
\epsilon_{\phi}^{(t)}
\!\! = \!\!
\max_{i\in[n]}\max_{m\in[k]}
\left|
\widehat{\Phi}_i^{(t)}(m)-\Phi_i^{(t)}(m)
\right|.
\label{eq:app_phi_error}
\end{equation}

\begin{lemma}[Stability of an approximate argmin]
\label{lemma:app_argmin_stability}
For a fixed sample and communication round, let $q_i^\star=\arg\min_m \Phi_i(m)$ and $\widehat{q}_i=\arg\min_m \widehat{\Phi}_i(m)$.
If $\max_m|\widehat{\Phi}_i(m)-\Phi_i(m)|\le \epsilon_i$, then
\begin{equation}
\Phi_i(\widehat{q}_i)\le \Phi_i(q_i^\star)+2\epsilon_i .
\label{eq:app_argmin_stability}
\end{equation}
\end{lemma}

\begin{proof}
The uniform error bound implies
\begin{equation}
\begin{aligned}
\!\!\Phi_i(m)\!\le \!\widehat{\Phi}_i(m)+\epsilon_i,
\widehat{\Phi}_i(m)\!\le\! \Phi_i(m)+\epsilon_i,
\quad \forall m\in[k].
\end{aligned}
\end{equation}
Applying the first inequality to $\widehat{q}_i$, using the optimality of $\widehat{q}_i$ under $\widehat{\Phi}_i$, and then applying the second inequality to $q_i^\star$ gives
\begin{equation}
\Phi_i(\widehat{q}_i)
\le \widehat{\Phi}_i(\widehat{q}_i)+\epsilon_i
\le \widehat{\Phi}_i(q_i^\star)+\epsilon_i
\le \Phi_i(q_i^\star)+2\epsilon_i .
\end{equation}
\end{proof}

\subsection{Effect of Proximal Term in \FedTriCD}
\label{sec:fedtricd_proximal}

We next relate local assignment drift to the proximal penalty and derive the margin induced by $\rho$.
Within communication round $t$, $m_i^{\mathrm{g}}$ denotes the cluster of sample $i$ in the broadcast assignment $\bF^{(t)}$.

\begin{lemma}[Frobenius distance and Hamming distance]
\label{lemma:app_frobenius_hamming}
For any two indicator matrices $\bF_h,\bF\in\mathcal{F}$, let
\begin{equation}
\Delta(\bF_h,\bF)
=
\#\{i\in[n]: r_i^{(h)}\neq m_i^{\mathrm{g}}\}
\end{equation}
be their row-wise Hamming distance, where $r_i^{(h)}$ is the cluster of sample $i$ in $\bF_h$ and $m_i^{\mathrm{g}}$ is its cluster in $\bF$.
Then
\begin{equation}
\|\bF_h-\bF\|_F^2=2\Delta(\bF_h,\bF).
\label{eq:app_frobenius_hamming}
\end{equation}
\end{lemma}

\begin{proof}
If the two assignments agree on row $i$, the row contribution to $\|\bF_h-\bF\|_F^2$ is zero.
If they disagree, the difference between the two one-hot row vectors contains one $1$ and one $-1$, and thus contributes $2$ to the squared Frobenius norm.
Summing over all rows gives \eqref{eq:app_frobenius_hamming}.
\end{proof}

\begin{lemma}[Local drift bound]
\label{lemma:app_local_drift}
At the beginning of communication round $t$, suppose client $h$ sets $\bF_h^{(t,0)}=\bF^{(t)}$.
Let $\bF_h^{(t,s)}$ be the local assignment after $s$ basic one-row update attempts within this round.
Then
\begin{equation}
\Delta(\bF_h^{(t,s)},\bF^{(t)})\le \min\{s,n\}.
\label{eq:app_local_drift}
\end{equation}
If each round runs $E$ local epochs and each epoch scans $n$ samples, then $s=nE$ and therefore $\Delta(\bF_h^{(t,nE)},\bF^{(t)})\le n$.
\end{lemma}

\begin{proof}
The claim is proved by induction on $s$.
For $s=0$, $\bF_h^{(t,0)}=\bF^{(t)}$, so the drift is zero.
During one basic row update attempt, at most one sample assignment can change, and hence the Hamming distance from $\bF^{(t)}$ can increase by at most one.
It is also always at most $n$ because there are only $n$ samples.
This proves \eqref{eq:app_local_drift}.
This is a worst-case combinatorial bound; the data-dependent effect of larger $E$ is absorbed into the surrogate error $\epsilon_{\phi}^{(t)}$.
\end{proof}

\begin{lemma}[Proximal margin for local moves]
\label{lemma:app_rho_stabilization}
For a feasible row update with all other rows fixed, if a non-reference candidate $m\neq m_i^{\mathrm{g}}$ beats the broadcast reference cluster under the proximal local objective, then
\begin{equation}
\psi(i,m_i^{\mathrm{g}},h)-\psi(i,m,h)>\rho .
\label{eq:app_rho_move_condition}
\end{equation}
\end{lemma}

\begin{proof}
By Lemma~\ref{lemma:app_frobenius_hamming}, $\frac{\rho}{2}\|\bF_h-\bF\|_F^2=\rho\Delta(\bF_h,\bF)$, so with all other rows fixed, the candidate-dependent proximal cost for row $i$ is $\rho\mathbb{I}\{m\neq m_i^{\mathrm{g}}\}$.
Thus, a non-reference candidate can beat the reference candidate only when
\begin{equation}
\psi(i,m,h)+\rho < \psi(i,m_i^{\mathrm{g}},h),
\end{equation}
which is equivalent to \eqref{eq:app_rho_move_condition}.
\end{proof}

\subsection{Server-Side Update Guarantee for \FedTriCD}
\label{sec:fedtricd_convergence}
At a given communication round, the local assignments $\{\bF_h^{(t,nE)}\}_{h=1}^H$ are fixed.
Here $m_i^{\mathrm{g}}$ is still the cluster of sample $i$ in the broadcast assignment $\bF^{(t)}$, which is the reference used by the proximal term throughout the round.
Because all clients use this same broadcast reference for the proximal penalty, the exact proximal aggregate score in Eq.~\eqref{eq:app_fedtricd_aggregate_scores} can be written as
\begin{equation}
P_i^{(t)}(m)
=
\sum_{h=1}^H \psi^{(t)}(i,m,h)
+
H\rho\,\mathbb{I}\{m\neq m_i^{\mathrm{g}}\}.
\label{eq:app_prox_aggregate}
\end{equation}
The server approximately minimizes $P_i^{(t)}(m)$ through the uploaded surrogate scores $\widehat{\Phi}_i^{(t)}(m)$.

\begin{theorem}[Server-side stability under bounded surrogate error]
\label{theorem:app_fedtricd_convergence}
Assume that empty clusters are forbidden, ties keep the broadcast reference assignment, and the server updates the global assignment row by row using the aggregated uploaded surrogate values.
Let $\widehat{q}_i$ be the server-selected cluster for sample $i$.
If there exists a round $t_0$ such that
\begin{equation}
2\epsilon_{\phi}^{(t)}<H\rho,
\qquad \forall t\ge t_0,
\label{eq:app_fedtricd_error_condition}
\end{equation}
then every nontrivial server-side row update after round $t_0$ achieves a strict decrease in the exact aggregate local objective after accounting for the proximal margin:
\begin{equation}
\sum_{h=1}^H \psi^{(t)}(i,\widehat{q}_i,h)
-
\sum_{h=1}^H \psi^{(t)}(i,m_i^{\mathrm{g}},h)
\le
2\epsilon_{\phi}^{(t)}-H\rho
<0.
\end{equation}
In particular, when the surrogate error vanishes asymptotically, i.e., $\epsilon_{\phi}^{(t)}\to 0$, condition \eqref{eq:app_fedtricd_error_condition} eventually holds.
\end{theorem}

\begin{proof}
For each row $i$, the broadcast reference cluster $m_i^{\mathrm{g}}$ has zero proximal penalty in \eqref{eq:app_prox_aggregate}.
Let $\widehat{q}_i$ be the cluster selected by the uploaded surrogate proximal values.
Since $P_i^{(t)}(m)=\Phi_i^{(t)}(m)$, Lemma~\ref{lemma:app_argmin_stability} gives
\begin{equation}
P_i^{(t)}(\widehat{q}_i)
\le
P_i^{(t)}(q_i^\star)+2\epsilon_{\phi}^{(t)}
\le
P_i^{(t)}(m_i^{\mathrm{g}})+2\epsilon_{\phi}^{(t)},
\end{equation}
where $q_i^\star=\arg\min_m P_i^{(t)}(m)$.
If the selected update is nontrivial, i.e., $\widehat{q}_i\neq m_i^{\mathrm{g}}$, then
\begin{equation}
\sum_{h=1}^H \psi^{(t)}(i,\widehat{q}_i,h)
-
\sum_{h=1}^H \psi^{(t)}(i,m_i^{\mathrm{g}},h)
\le
2\epsilon_{\phi}^{(t)}-H\rho
<0.
\end{equation}
If $\epsilon_{\phi}^{(t)}\to0$, then \eqref{eq:app_fedtricd_error_condition} holds for all sufficiently large $t$ because $\rho>0$.
\end{proof}

\noindent\textbf{Remarks.}
The theorem highlights the roles of the two key parameters in \FedTriCD.
The local period $E$ determines the number of local epochs performed before the next communication round and influences $\epsilon_{\phi}^{(t)}$ through the resulting local states and bound tightness.
The proximal coefficient $\rho$ discourages local moves that disagree with the global assignment unless their local objective improvement exceeds $\rho$, as shown in Lemma~\ref{lemma:app_rho_stabilization}.
The server-side guarantee holds once the resulting surrogate error is smaller than the proximal margin.

\begin{table}[t]
\centering
\caption{An overview of the datasets.}
\vspace{-1em}
\label{tab:Dataset Table}
\resizebox{0.82\columnwidth}{!}{
\begin{tabular}{cccc}
\toprule
No. & Datasets                & Instances & Features \\ 
\midrule
1 & \YearPredictionMSD~\cite{Bera19Fair} & 515,345    & 90       \\
2 & \Epileptic~\cite{Epileptic} & 11,500      & 179       \\
3 & \NYC~\cite{NYC}               & 93,817      & 100        \\
4 & \Kegg~\cite{shu2022Ball}               & 65,554    & 28       \\
5 & \Crime~\cite{crime}  & 1,993     & 127       \\
6 & \Shuttle~\cite{moreno2012study} & 57,999     & 9       \\
7 & \Letter~\cite{moreno2012study}     & 20,000      & 16       \\
8 & \Mnist~\cite{lecun2002gradient}  & 70,000       & 784       \\ 
9 & \Cifarten~\cite{krizhevsky2009learning}  & 60,000       & 3072       \\ 
10 & \Cifarhundred~\cite{krizhevsky2009learning}  & 60,000       & 3072       \\ 
\bottomrule
\end{tabular}
}
\vspace{-1.3em}
\end{table}

\begin{table*}[ht]
\caption{Running time (s) and speedup over \Lloyd for different $k$-means algorithms on multiple datasets. For each dataset and $k$, bold values indicate the best result.}
\vspace{-0.8em}
\label{tab:timespeedup}
\resizebox{\textwidth}{!}{
\begin{tabular}{cccccccccccccccc}
\toprule
\multirow{2}{*}{Dataset} & \multirow{2}{*}{k} & \multicolumn{7}{c}{Time} & \multicolumn{7}{c}{Speed} \\
\cmidrule(lr){3-9}\cmidrule(lr){10-16}
& & \Lloyd & \CDKM & \Hamerly & \Yinyang & \ballkmeans & \LBKC & \TriCD & \Lloyd & \CDKM & \Hamerly & \Yinyang & \ballkmeans & \LBKC & \TriCD \\
\midrule
\multirow{3}{*}{\YearPredictionMSD}
& 10 & 1.28 & 1.56 & 0.82 & 0.83 & 1.30 & 1.04 & \textbf{0.20} & 1.00 & 0.82 & 1.55 & 1.54 & 0.98 & 1.23 & \textbf{6.34} \\
& 30 & 3.03 & 5.77 & 1.74 & 1.75 & 1.30 & 1.14 & \textbf{0.69} & 1.00 & 0.53 & 1.74 & 1.73 & 2.34 & 2.66 & \textbf{4.38} \\
& 50 & 3.21 & 8.54 & 2.35 & 2.44 & 0.98 & \textbf{0.86} & 0.94 & 1.00 & 0.38 & 1.37 & 1.32 & 3.27 & \textbf{3.74} & 3.42 \\
\midrule
\multirow{3}{*}{\Epileptic}
& 10 & 1.99 & 5.16 & 2.11 & 2.02 & 2.17 & 1.82 & \textbf{0.55} & 1.00 & 0.39 & 0.94 & 0.99 & 0.91 & 1.09 & \textbf{3.62} \\
& 30 & 3.71 & 5.17 & 2.57 & 2.65 & 1.41 & 1.29 & \textbf{0.64} & 1.00 & 0.72 & 1.44 & 1.40 & 2.63 & 2.86 & \textbf{5.77} \\
& 50 & 10.47 & 14.50 & 7.68 & 8.05 & 2.61 & 2.47 & \textbf{1.50} & 1.00 & 0.72 & 1.36 & 1.30 & 4.02 & 4.24 & \textbf{7.00} \\
\midrule
\multirow{3}{*}{\NYC}
& 10 & 5.46 & 8.56 & 4.12 & 4.25 & 4.70 & 3.89 & \textbf{1.12} & 1.00 & 0.64 & 1.33 & 1.28 & 1.16 & 1.40 & \textbf{4.87} \\
& 30 & 34.67 & 31.34 & 17.34 & 18.60 & 8.71 & 11.89 & \textbf{3.92} & 1.00 & 1.11 & 2.00 & 1.86 & 3.98 & 2.92 & \textbf{8.85} \\
& 50 & 77.37 & 42.29 & 35.93 & 39.12 & 17.88 & 20.44 & \textbf{5.19} & 1.00 & 1.83 & 2.15 & 1.98 & 4.33 & 3.79 & \textbf{14.92} \\
\midrule
\multirow{3}{*}{\Kegg}
& 10 & 2.10 & 7.19 & 1.27 & 1.45 & 1.24 & 1.04 & \textbf{0.85} & 1.00 & 0.29 & 1.65 & 1.45 & 1.69 & 2.02 & \textbf{2.46} \\
& 30 & 20.20 & 13.00 & 7.21 & 8.63 & 5.06 & 4.62 & \textbf{1.78} & 1.00 & 1.55 & 2.80 & 2.34 & 3.99 & 4.37 & \textbf{11.34} \\
& 50 & 20.60 & 25.67 & 7.67 & 9.23 & \textbf{3.58} & 4.21 & 3.61 & 1.00 & 0.80 & 2.69 & 2.23 & \textbf{5.76} & 4.89 & 5.71 \\
\midrule
\multirow{3}{*}{\Crime}
& 10 & 0.08 & 0.26 & 0.03 & \textbf{0.03} & 0.06 & 0.07 & 0.03 & 1.00 & 0.32 & 2.50 & \textbf{3.10} & 1.37 & 1.17 & 3.06 \\
& 30 & 0.18 & 0.53 & 0.12 & 0.12 & 0.13 & \textbf{0.11} & 0.12 & 1.00 & 0.35 & 1.48 & 1.57 & 1.41 & \textbf{1.63} & 1.59 \\
& 50 & 0.31 & 0.48 & 0.18 & 0.17 & 0.11 & 0.11 & \textbf{0.07} & 1.00 & 0.65 & 1.73 & 1.81 & 2.88 & 2.92 & \textbf{4.36} \\
\midrule
\multirow{3}{*}{\Shuttle}
& 10 & 0.72 & 1.74 & 0.30 & 0.42 & 0.35 & 0.31 & \textbf{0.27}
     & 1.00 & 0.41 & 2.35 & 1.70 & 2.07 & 2.30 & \textbf{2.69} \\
& 30 & 3.83 & 3.43 & 0.87 & 1.63 & 0.61 & 0.98 & \textbf{0.43}
     & 1.00 & 1.12 & 4.42 & 2.36 & 6.25 & 3.92 & \textbf{8.91} \\
& 50 & 3.44 & 17.20 & 0.79 & 1.82 & \textbf{0.39} & 0.72 & 1.58
     & 1.00 & 0.20 & 4.33 & 1.89 & \textbf{8.78} & 4.76 & 2.18 \\
\midrule
\multirow{3}{*}{\Letter}
& 10 & 0.48 & 1.43 & 0.20 & 0.23 & 0.30 & 0.31 & \textbf{0.17}
     & 1.00 & 0.34 & 2.39 & 2.15 & 1.60 & 1.54 & \textbf{2.86} \\
& 30 & 4.52 & 4.85 & 1.24 & 1.43 & 1.30 & 1.52 & \textbf{0.47}
     & 1.00 & 0.93 & 3.66 & 3.16 & 3.48 & 2.98 & \textbf{9.54} \\
& 50 & 4.58 & 8.07 & 1.48 & 1.83 & 1.30 & 1.16 & \textbf{0.96}
     & 1.00 & 0.57 & 3.10 & 2.51 & 3.52 & 3.95 & \textbf{4.79} \\
\midrule
\multirow{3}{*}{\Mnist}
& 10 & 93.07 & 23.51 & 54.65 & 53.96 & 57.83 & 45.64 & \textbf{4.51}
     & 1.00 & 3.96 & 1.70 & 1.72 & 1.61 & 2.04 & \textbf{20.64} \\
& 30 & 188.60 & 94.77 & 87.54 & 87.03 & 51.01 & 44.78 & \textbf{9.92}
     & 1.00 & 1.99 & 2.15 & 2.17 & 3.70 & 4.21 & \textbf{19.01} \\
& 50 & 522.26 & 124.05 & 219.48 & 220.50 & 97.69 & 84.68 & \textbf{10.43}
     & 1.00 & 4.21 & 2.38 & 2.37 & 5.35 & 6.17 & \textbf{50.07} \\
\midrule
\multirow{3}{*}{\Cifarten}
& 10 & 904.74 & 225.48 & 313.49 & 296.61 & 260.55 & 215.29 & \textbf{44.72}
     & 1.00 & 4.01 & 2.89 & 3.05 & 3.47 & 4.20 & \textbf{20.23} \\
& 30 & 3328.03 & 962.22 & 1084.41 & 1076.13 & 386.43 & 299.37 & \textbf{90.34}
     & 1.00 & 3.46 & 3.07 & 3.09 & 8.61 & 11.12 & \textbf{36.84} \\
& 50 & 5148.17 & 1930.39 & 1659.50 & 1663.13 & 330.96 & 285.45 & \textbf{144.59}
     & 1.00 & 2.67 & 3.10 & 3.10 & 15.56 & 18.04 & \textbf{35.61} \\
\midrule
\multirow{3}{*}{\Cifarhundred}
& 10 & 1056.10 & 200.93 & 366.78 & 356.31 & 307.92 & 249.80 & \textbf{42.41}
     & 1.00 & 5.26 & 2.88 & 2.96 & 3.43 & 4.23 & \textbf{24.90} \\
& 30 & 3344.64 & 1558.35 & 1085.03 & 1077.91 & 372.61 & 299.70 & \textbf{132.18}
     & 1.00 & 2.15 & 3.08 & 3.10 & 8.98 & 11.16 & \textbf{25.30} \\
& 50 & 3173.56 & 2403.16 & 1095.03 & 1095.15 & 192.34 & 173.63 & \textbf{131.76}
     & 1.00 & 1.32 & 2.90 & 2.90 & 16.50 & 18.28 & \textbf{24.09} \\
\bottomrule
\vspace{-2.5em}
\end{tabular}}
\end{table*}

\begin{table*}[ht]
\caption{Dominant candidate evaluations and reduction factors relative to \Lloyd across datasets and $k$ values.}
\vspace{-0.8em}
\label{tab:distancespeedup}
\resizebox{\textwidth}{!}{
\begin{tabular}{lccccccccccccccc}
\toprule
\multirow{2}{*}{Dataset} & \multirow{2}{*}{k} & \multicolumn{7}{c}{Evaluations} & \multicolumn{7}{c}{Reduction Factor} \\
\cmidrule(lr){3-9}\cmidrule(lr){10-16}
& & \Lloyd & \CDKM & \Hamerly & \Yinyang & \ballkmeans & \LBKC & \TriCD & \Lloyd & \CDKM & \Hamerly & \Yinyang & \ballkmeans & \LBKC & \TriCD \\
\midrule
\multirow{3}{*}{\YearPredictionMSD}
& 10 & 1.E+07 & 1.E+07 & 5.E+06 & 4.E+06 & 1.E+07 & 9.E+06 & \textbf{1.E+06}
& 1.00 & 1.00 & 2.05 & 2.29 & 1.03 & 1.09 & \textbf{10.57} \\
& 30 & 3.E+07 & 3.E+07 & 1.E+07 & 1.E+07 & 3.E+07 & 2.E+07 & \textbf{3.E+06}
& 1.00 & 1.00 & 1.95 & 2.06 & 1.03 & 1.18 & \textbf{10.11} \\
& 50 & 3.E+07 & 3.E+07 & 2.E+07 & 2.E+07 & 3.E+07 & 3.E+07 & \textbf{4.E+06}
& 1.00 & 1.00 & 1.41 & 1.47 & 1.05 & 1.23 & \textbf{8.40} \\
\midrule
\multirow{3}{*}{\Epileptic}
& 10 & 8.E+06 & 8.E+06 & 5.E+06 & 4.E+06 & 8.E+06 & 5.E+06 & \textbf{1.E+06}
& 1.00 & 1.00 & 1.63 & 1.82 & 1.02 & 1.75 & \textbf{6.45} \\
& 30 & 2.E+07 & 2.E+07 & 8.E+06 & 8.E+06 & 2.E+07 & 6.E+06 & \textbf{2.E+06}
& 1.00 & 1.00 & 1.93 & 2.08 & 1.00 & 2.45 & \textbf{9.88} \\
& 50 & 5.E+07 & 5.E+07 & 3.E+07 & 3.E+07 & 5.E+07 & 2.E+07 & \textbf{3.E+06}
& 1.00 & 1.00 & 1.60 & 1.67 & 1.00 & 2.73 & \textbf{14.47} \\
\midrule
\multirow{3}{*}{\NYC}
& 10 & 3.E+07 & 3.E+07 & 1.E+07 & 1.E+07 & 2.E+07 & 2.E+07 & \textbf{4.E+06}
& 1.00 & 1.00 & 2.43 & 2.40 & 1.36 & 1.94 & \textbf{6.79} \\
& 30 & 2.E+08 & 2.E+08 & 7.E+07 & 7.E+07 & 1.E+08 & 9.E+07 & \textbf{1.E+07}
& 1.00 & 1.00 & 3.61 & 3.36 & 1.96 & 2.82 & \textbf{24.38} \\
& 50 & 6.E+08 & 6.E+08 & 2.E+08 & 2.E+08 & 4.E+08 & 2.E+08 & \textbf{1.E+07}
& 1.00 & 1.00 & 3.29 & 2.97 & 1.50 & 3.32 & \textbf{44.51} \\
\midrule
\multirow{3}{*}{\Kegg}
& 10 & 2.E+07 & 2.E+07 & 8.E+06 & 8.E+06 & 2.E+07 & 1.E+07 & \textbf{4.E+06}
& 1.00 & 1.00 & 2.86 & 3.08 & 1.41 & 1.68 & \textbf{6.49} \\
& 30 & 2.E+08 & 2.E+08 & 5.E+07 & 6.E+07 & 2.E+08 & 9.E+07 & \textbf{6.E+06}
& 1.00 & 1.00 & 4.26 & 3.88 & 1.30 & 2.73 & \textbf{37.96} \\
& 50 & 3.E+08 & 3.E+08 & 8.E+07 & 9.E+07 & 2.E+08 & 9.E+07 & \textbf{9.E+06}
& 1.00 & 1.00 & 3.39 & 3.07 & 1.54 & 3.04 & \textbf{28.53} \\
\midrule
\multirow{3}{*}{\Crime}
& 10 & 4.E+05 & 4.E+05 & 2.E+05 & \textbf{1.E+05} & 4.E+05 & 3.E+05 & 2.E+05
& 1.00 & 1.00 & 2.60 & \textbf{3.20} & 1.00 & 1.47 & 2.53 \\
& 30 & 2.E+06 & 2.E+06 & 1.E+06 & 9.E+05 & 2.E+06 & 1.E+06 & \textbf{5.E+05}
& 1.00 & 1.00 & 1.52 & 1.68 & 0.83 & 1.61 & \textbf{2.93} \\
& 50 & 3.E+06 & 3.E+06 & 2.E+06 & 1.E+06 & 2.E+06 & 1.E+06 & \textbf{6.E+05}
& 1.00 & 1.00 & 1.73 & 1.92 & 1.13 & 1.85 & \textbf{4.73} \\
\midrule
\multirow{3}{*}{\Shuttle}
& 10 & 1.E+07 & 8.E+06 & 4.E+06 & 4.E+06 & 5.E+06 & 6.E+06 & \textbf{2.E+06}
& 1.00 & 1.50 & 3.40 & 2.80 & 2.30 & 2.13 & \textbf{7.86} \\
& 30 & 8.E+07 & 2.E+07 & 2.E+07 & 2.E+07 & 2.E+07 & 2.E+07 & \textbf{3.E+06}
& 1.00 & 4.08 & 5.14 & 3.27 & 3.48 & 4.61 & \textbf{24.87} \\
& 50 & 7.E+07 & 1.E+08 & 2.E+07 & 3.E+07 & 2.E+07 & 1.E+07 & \textbf{7.E+06}
& 1.00 & 0.71 & 4.06 & 2.29 & 3.13 & 6.27 & \textbf{10.08} \\
\midrule
\multirow{3}{*}{\Letter}
& 10 & 8.E+06 & 6.E+06 & 3.E+06 & 2.E+06 & 7.E+06 & 8.E+06 & \textbf{1.E+06}
& 1.00 & 1.19 & 2.71 & 3.16 & 1.03 & 0.97 & \textbf{5.73} \\
& 30 & 8.E+07 & 3.E+07 & 2.E+07 & 2.E+07 & 8.E+07 & 8.E+07 & \textbf{3.E+06}
& 1.00 & 3.18 & 4.42 & 4.68 & 1.06 & 1.11 & \textbf{30.29} \\
& 50 & 9.E+07 & 5.E+07 & 3.E+07 & 3.E+07 & 1.E+08 & 7.E+07 & \textbf{4.E+06}
& 1.00 & 1.91 & 3.16 & 3.29 & 0.81 & 1.19 & \textbf{19.92} \\
\midrule
\multirow{3}{*}{\Mnist}
& 10 & 8.E+07 & 3.E+07 & 2.E+07 & 2.E+07 & 8.E+07 & 8.E+07 & \textbf{5.E+06}
& 1.00 & 2.63 & 4.41 & 5.01 & 1.00 & 0.97 & \textbf{14.16} \\
& 30 & 2.E+08 & 2.E+08 & 6.E+07 & 6.E+07 & 2.E+08 & 1.E+08 & \textbf{1.E+07}
& 1.00 & 1.39 & 3.56 & 3.81 & 1.00 & 1.58 & \textbf{17.82} \\
& 50 & 7.E+08 & 2.E+08 & 2.E+08 & 2.E+08 & 7.E+08 & 3.E+08 & \textbf{1.E+07}
& 1.00 & 3.06 & 3.40 & 3.53 & 1.00 & 2.25 & \textbf{47.51} \\
\midrule
\multirow{3}{*}{\Cifarten}
& 10 & 9.E+07 & 4.E+07 & 1.E+07 & 1.E+07 & 9.E+07 & 8.E+07 & \textbf{6.E+06}
& 1.00 & 2.46 & 6.29 & 7.14 & 1.00 & 1.05 & \textbf{14.29} \\
& 30 & 4.E+08 & 2.E+08 & 9.E+07 & 9.E+07 & 4.E+08 & 3.E+08 & \textbf{1.E+07}
& 1.00 & 1.96 & 3.95 & 4.16 & 1.00 & 1.19 & \textbf{24.39} \\
& 50 & 6.E+08 & 4.E+08 & 2.E+08 & 2.E+08 & 6.E+08 & 5.E+08 & \textbf{3.E+07}
& 1.00 & 1.51 & 3.58 & 3.73 & 1.00 & 1.23 & \textbf{22.46} \\
\midrule
\multirow{3}{*}{\Cifarhundred}
& 10 & 1.E+08 & 3.E+07 & 2.E+07 & 1.E+07 & 1.E+08 & 1.E+08 & \textbf{6.E+06}
& 1.00 & 3.13 & 6.10 & 6.89 & 1.00 & 1.07 & \textbf{17.76} \\
& 30 & 4.E+08 & 3.E+08 & 9.E+07 & 9.E+07 & 4.E+08 & 3.E+08 & \textbf{2.E+07}
& 1.00 & 1.23 & 4.00 & 4.22 & 1.00 & 1.21 & \textbf{18.06} \\
& 50 & 3.E+08 & 5.E+08 & 1.E+08 & 1.E+08 & 3.E+08 & 3.E+08 & \textbf{2.E+07}
& 1.00 & 0.74 & 3.20 & 3.36 & 1.00 & 1.28 & \textbf{14.98} \\

\bottomrule

\vspace{-3em}
\end{tabular}}
\end{table*}

\begin{figure*}[t]
  \centering
  \includegraphics[width=\linewidth]{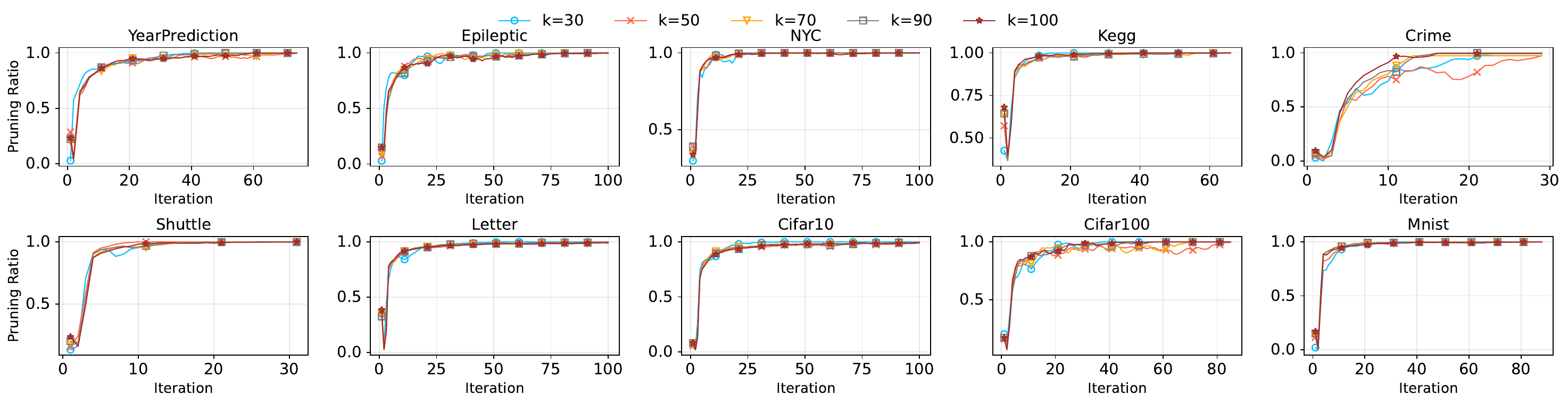}
  \vspace{-2.3em}
  \caption{Pruning ratio of \TriCD over iterations under different values of $k$.}
\label{fig:PruningRatio}
  \vspace{-1.2em}
\end{figure*}
\begin{table}[t]
  \centering
  \caption{Comparison of communication rounds in VFL methods.}
  \vspace{-0.8em}
  \label{tab:cost_round_compare}
  \scriptsize
  \resizebox{\columnwidth}{!}{
  \begin{tabular}{@{}lcccccccccc@{}}
    \toprule
    Method 
    & \YearPredictionMSD & \Epileptic & \NYC & \Kegg & \Crime 
    & \Shuttle & \Letter & \Mnist & \Cifarten & \Cifarhundred \\
    \midrule
    \FFFKM    
    & 4.E+02 & 4.E+02 & 4.E+02 & 4.E+02 & 2.E+01 
    & 4.E+02 & 1.E+03 & 1.E+03 & 5.E+02 & 5.E+02 \\
    \FedTriCD 
    & 2.E+01 & 3.E+01 & 1.E+01 & 2.E+01 & 2.E+01 
    & 2.E+01 & 3.E+01 & 2.E+01 & 2.E+01 & 5.E+01 \\
    \bottomrule
  \end{tabular}
  }
  \vspace{-1.2em}
\end{table}

\section{EXPERIMENTS}
\label{sec:experiments}

We evaluate the proposed methods in terms of computational efficiency, communication rounds, and clustering quality on a variety of benchmark datasets, and compare them with state-of-the-art baselines.
To ensure reproducibility, we provide the open-source code used in our experiments\footnote{\textcolor{black}{\href{https://github.com/whu-totemdb/FedTriCD.git}{https://github.com/whu-totemdb/FedTriCD.git}}}.

\subsection{Settings}
\myparagraph{Datasets}
We evaluate \FedTriCD and other comparable methods on widely used public benchmark datasets, including \YearPredictionMSD~\cite{Bera19Fair}, \Epileptic~\cite{Epileptic}, \NYC~\cite{NYC}, \Kegg~\cite{shu2022Ball}, \Crime~\cite{crime}, \Shuttle~\cite{moreno2012study}, \Letter~\cite{moreno2012study}, \Mnist~\cite{lecun2002gradient}, \Cifarten~\cite{krizhevsky2009learning}, \Cifarhundred~\cite{krizhevsky2009learning}.
These datasets have been commonly adopted in prior clustering studies and recent federated clustering benchmarks~\cite{Ding16KMeans,li2022differentially,Bera19Fair,shu2022Ball}.
For each dataset, we use the full set of available instances and features in the experiments.
Table~\ref{tab:Dataset Table} summarizes the dataset statistics.

\begin{figure*}[t]
  \centering
  \includegraphics[width=\linewidth]{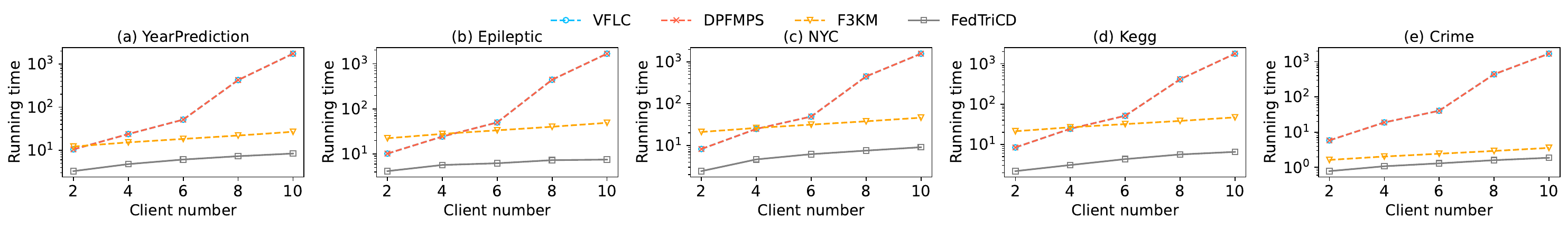}
  \vspace{-2.0em}
  \caption{Running time with varying numbers of clients.}
\label{fig:varclient}
  \vspace{-1.3em}
\end{figure*}
\begin{figure*}[t]
  \centering
  \includegraphics[width=\linewidth]{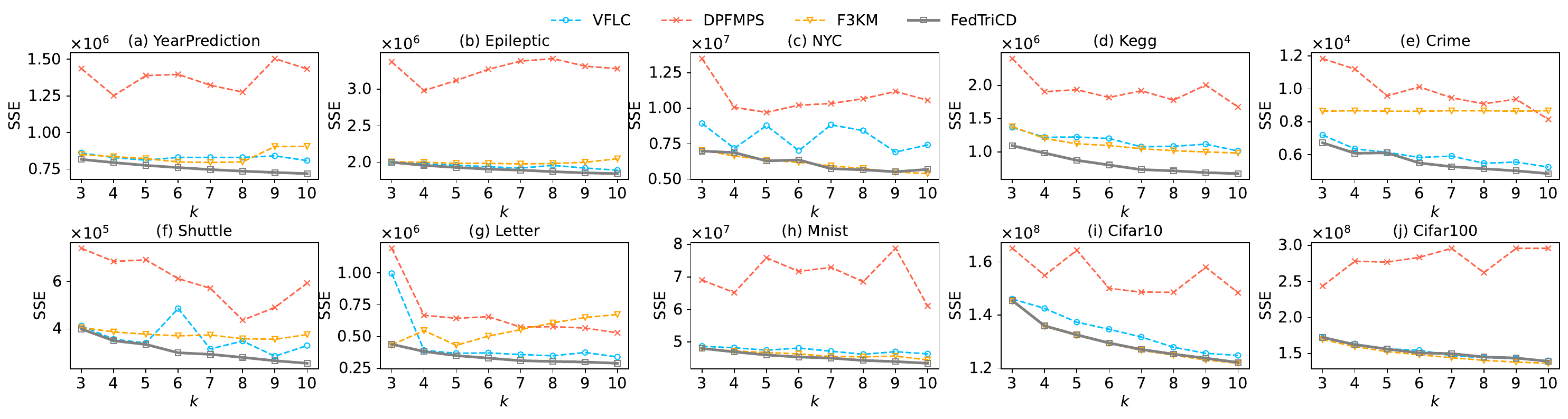}
  \vspace{-2.6em}
  \caption{SSE of different methods across datasets for varying values of $k$.}
   \vspace{-1.7em}
  \label{fig:SSE}
\end{figure*}

\myparagraph{Implementations}
All experiments were carried out on a Windows 11 Professional system, equipped with a 24-core Intel i9-14900KF CPU, 128 GB of RAM, and a 2 TB SSD. 
The code was implemented in MATLAB R2022a.
For the baseline methods, we adopted the parameter settings recommended in their original papers.
To ensure a fair comparison, all methods use the same initialization.
For the \Yinyang algorithm, we set the number of center groups to $t=k/10$, following common practice.
Unless otherwise specified, we use four clients and evenly partition the features among them to construct a controlled VFL setting.

\myparagraph{Baselines}
We compare the proposed methods with nine baselines: \Lloyd, \CDKM~\cite{nie2022CDKM}, \Hamerly~\cite{Hamerly10Making}, \Yinyang~\cite{ding15Yinyang}, \ballkmeans~\cite{shu2022Ball}, \LBKC~\cite{zhang2024speeding}, \VFLC~\cite{Ding16KMeans}, \DPFMPS~\cite{li2022differentially}, and \FFFKM~\cite{Zhu23F3KM}.
\Lloyd and \CDKM serve as unaccelerated centralized baselines; \Hamerly, \Yinyang, \ballkmeans, and \LBKC are representative fast centralized methods; and \VFLC, \DPFMPS, and \FFFKM are VFL clustering baselines.
The clustering iterations are terminated when the change in SSE is smaller than  $1 \times 10^{-5}$.

\myparagraph{Metrics}
We evaluate the methods along three dimensions: computation, clustering quality, and communication rounds.
We report running time and the number of dominant candidate evaluations.
For Lloyd-type methods (\Lloyd, \Hamerly, \Yinyang, \ballkmeans, \LBKC), the evaluation count refers to point-to-center Euclidean distance computations; 
for \CDKM and \TriCD, it refers to the inner-product evaluations of $s_{im}$. 
This count provides an implementation-agnostic measure of the main computational burden and directly reflects the effect of pruning.
We evaluate clustering quality using the sum of squared errors (SSE), as introduced in Section~\ref{sec:pre}.
For VFL methods, we report the number of communication rounds. 
All $k$-means-based methods are randomly initialized, with the same initialization shared across methods in each run.
For each dataset and each parameter setting, we run every method five times with different random initializations and report the average results.

\begin{figure*}[t]
  \centering
  \includegraphics[width=\linewidth]{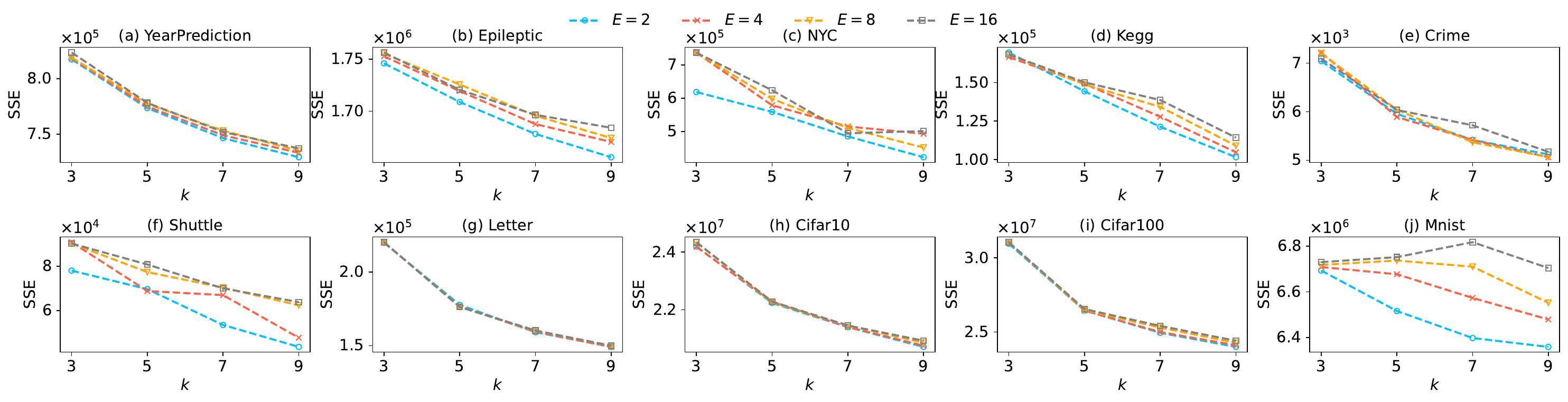}
  \vspace{-2.4em}
  \caption{SSE of \FedTriCD under different values of $E$.}
\label{fig:varE}
  \vspace{-1.5em}
\end{figure*}
\begin{figure*}[t]
  \centering
  \includegraphics[width=\linewidth]{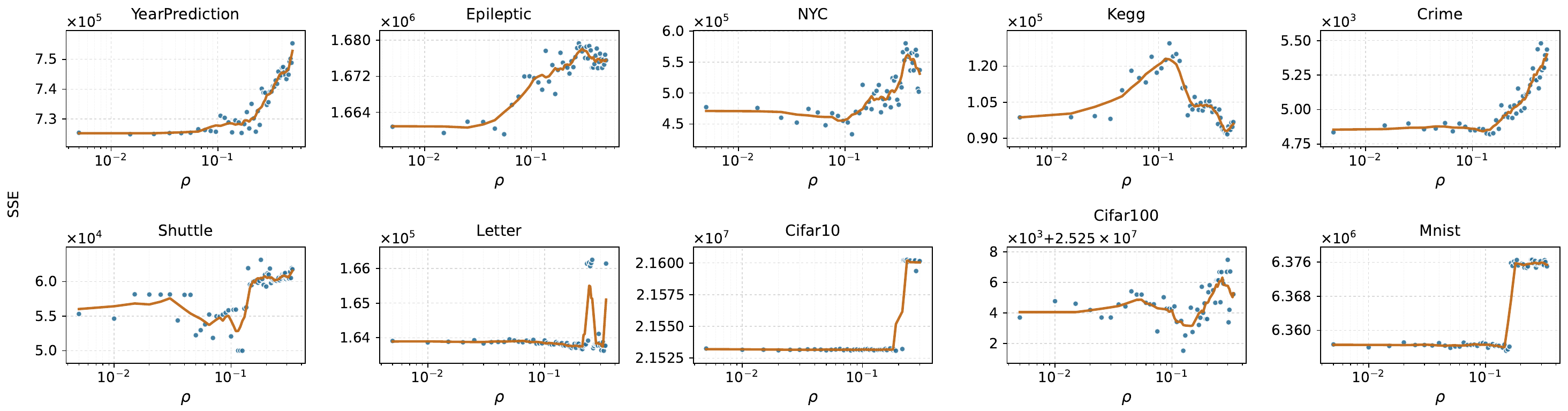}
  \vspace{-2.4em}
  \caption{SSE of \FedTriCD under different values of $\rho$.}
\label{fig:varrho}
  \vspace{-1.4em}
\end{figure*}
\begin{figure*}[t]
  \centering
  \includegraphics[width=\linewidth]{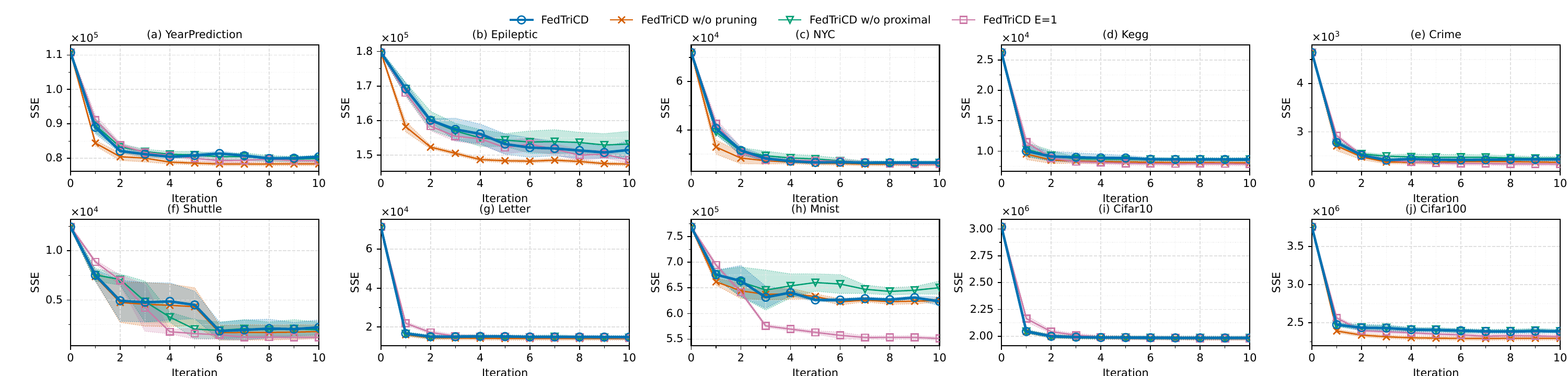}
  \vspace{-2.2em}
  \caption{Ablation study of \FedTriCD.}
\label{fig:ablation}
  \vspace{-1.8em}
\end{figure*}

\subsection{Efficiency of Proposed Methods}
\myparagraph{Comparison of Computational Efficiency} 
We evaluate computational efficiency from two complementary perspectives: runtime and the number of dominant candidate calculations. 
Specifically, Table~\ref{tab:timespeedup} reports the running time and the corresponding speedup over \Lloyd, while Table~\ref{tab:distancespeedup} reports the number of objective calculations and the corresponding reduction factor relative to \Lloyd. 

As shown in Table~\ref{tab:timespeedup}, \TriCD achieves the best computational efficiency in most settings, delivering speedups over \Lloyd ranging from $1.59\times$ to $50.07\times$ and averaging about $12.5\times$.
The advantage is especially clear on large or high-dimensional datasets such as \Mnist, \Cifarten, and \Cifarhundred. 
Table~\ref{tab:distancespeedup} shows that this gain mainly comes from pruning expensive objective evaluations: \TriCD attains the smallest number of objective calculations in 29 out of 30 settings, with an average reduction factor of about $16.96\times$ relative to \Lloyd. 
Moreover, \TriCD consistently improves over \CDKM\ in runtime, indicating that the performance gain mainly comes from the proposed pruning strategy.


We further investigate how the pruning effectiveness evolves over clustering iterations. 
Fig.~\ref{fig:PruningRatio} shows how the pruning ratio of \TriCD evolves over iterations on different datasets, measured as the fraction of inner product evaluations avoided.
Across all datasets, the pruned fraction increases sharply after only a few iterations and then stabilizes. 
This pattern suggests that many assignments stabilize early, and \TriCD uses lower bounds to avoid many exact inner-product evaluations.

\myparagraph{Comparison of Communication Rounds}
Table~\ref{tab:cost_round_compare} reports the communication rounds of iterative VFL baselines and \FedTriCD; one-shot methods are discussed separately because they communicate once but incur different server-side costs.
\VFLC and \DPFMPS communicate only once via one-shot summaries, but their server-side matching cost can scale as $\mathcal{O}(k^H)$ with the number of clients $H$, which quickly becomes prohibitive (Fig.~\ref{fig:varclient}). 
In contrast, \FFFKM relies on iterative optimization and thus requires substantially more rounds as the problem scale increases.
As the dataset scale increases, \FFFKM requires tens to thousands of communication rounds across datasets, leading to high round latency and reduced efficiency in practical deployments.
By contrast, \FedTriCD requires only a small number of communication rounds across all datasets, indicating that its collaborative update mechanism effectively controls the number of communication rounds as the problem scale grows.

\myparagraph{Comparison of Clustering Quality}
Fig.~\ref{fig:SSE} compares the SSE values of different methods across datasets and values of $k$.
The results show that \FedTriCD maintains competitive clustering quality while reducing local computation and communication rounds.
Together with Fig.~\ref{fig:SSE}, these results indicate that the proposed method improves efficiency without causing substantial degradation in SSE.

\myparagraph{Impact of Communication Period $E$}
We vary $E \in \{2,4,8,16\}$ and report the resulting SSE (Fig.~\ref{fig:varE}). 
Across datasets, \FedTriCD is generally robust to $E$, and the SSE curves largely overlap for moderate $E$ (e.g., $E=4$ or $8$). 
When $E$ becomes large, a slight increase in SSE can be observed on some datasets, which is expected due to the use of increasingly stale global assignment information during local updates.
In our experiments, $E=4$ provides a good trade-off.

\myparagraph{Impact of Proximal Coefficient $\rho$}
We investigate the impact of the proximal coefficient $\rho$ by fixing the number of local epochs to $E=2$ and the number of clusters to $k=4$, while varying $\rho$. 
The SSE of \FedTriCD across different datasets is reported in Fig.~\ref{fig:varrho}.
When $\rho$ is too small, local assignments can drift away from the global consensus, degrading SSE.
When $\rho$ is too large, local updates become overly constrained and may suppress beneficial reassignments.
Moderate values of $\rho$ provide a better balance by stabilizing local updates without suppressing useful reassignment.

\myparagraph{Ablation Study}
Fig.~\ref{fig:ablation} reports the SSE of \FedTriCD and its ablated variants.
Across all datasets, \FedTriCD quickly reduces the SSE in the first few iterations and then reaches a stable solution.
Compared with the variant without pruning, \FedTriCD achieves comparable SSE on most datasets.
This indicates that pruning reduces redundant computations without causing noticeable degradation in clustering quality.
Removing the proximal term leads to unstable trajectories across several datasets, indicating that proximal regularization helps control client drift during multiple local updates.
The $E=1$ variant often converges more slowly since each communication round performs only one iteration of local updates.
On some datasets, however, more frequent communication exploits fresher global information, leading to even lower SSE.



\section{Conclusion}
This paper investigated fast $k$-means in collaborative settings where multiple parties cannot share raw data.
To address this challenge, we proposed \TriCD, an exact pruning method for \CDKM that builds on-demand bounds with negligible extra space, and further developed \FedTriCD, an efficient VFL $k$-means algorithm that avoids explicitly materializing centroids and leverages triangle-inequality-based pruning to reduce redundant computations.
Our theoretical analysis proved the exactness and finite convergence of \TriCD and established a bounded-error server-side update guarantee for \FedTriCD.
Experiments showed that \TriCD achieves up to an $50.07\times$ speedup over \Lloyd's algorithm, while \FedTriCD achieves competitive clustering quality with fewer local evaluations and communication rounds than existing VFL baselines.

Future work will extend \FedTriCD to other clustering objectives and more heterogeneous VFL settings, such as imbalanced feature partitions and constrained clustering.
We will also investigate how to handle streaming data.

\section{Acknowledgment}
This work was supported by the National Key R$\&$D Program of China (2023YFB4503600), National Natural Science Foundation of China (62202338 and 62372337), the Key R$\&$D Program of Hubei Province (2023BAB081), and the Hubei Key Laboratory of Big Data in Science and Technology (No.~E6KF001001).

\section{AI-Generated Content Acknowledgement}
We used ChatGPT only to check grammar in the manuscript. The authors completed all technical ideas, algorithmic designs, experimental analyses, manuscript writing, and code.

\bibliographystyle{IEEEtran}

\bibliography{sample-base}

\end{document}